\documentclass{aastex}          
\usepackage{spr-astr-addons}    
\usepackage{url}
\usepackage{amsmath, amsfonts}
\usepackage{graphicx}
\usepackage{epstopdf}
\usepackage{tabularx}

\begin{document}
\title{Modeling of non-rotating neutron stars in \\ minimal dilatonic gravity}

\shorttitle{<Short article title>}
\shortauthors{<Autors et al.>}

\author{P.~Fiziev\altaffilmark{1,}\altaffilmark{3}}
\altaffiltext{1}{JINR, Dubna, 141980 Moscow Region, Russia}
\and
\author{K.~Marinov\altaffilmark{2,}\altaffilmark{1}}
\email{marinov.kalin@gmail.com} 

\altaffiltext{2}{Institute for Nuclear Research and Nuclear Energy, Bulgarian
Academy of Sciences, bul. Tzarigradsko chaussee 72, Sofia 1784, Bulgaria}
\altaffiltext{3}{Sofia University Foundation for Theoretical and Computational
Physics and Astrophysics, 5 James Bourchier Blvd., 1164 Sofia, Bulgaria}
\begin{abstract}

The model of minimal dilatonic gravity (MDG), called also the massive Branse-Dicke model with $\omega =0$,
is an alternative model of gravitation, which uses one Branse-Dicke gravitation-dilaton field $\Phi$
and offers a simultaneous explanation of the effects of dark energy (DE) and dark matter (DM).
Here we present an extensive research of non-rotating neutron star models in MDG
with four different realistic equations of state (EOS), which are in agreement with the latest observational data.
The equations describing static spherically symmetric stars in MDG are solved numerically.
The effects corresponding to DE and DM are clearly seen and discussed.

\end{abstract}


%

\section{Introduction}\label{intro}

It is well known that General Relativity (GR) and the Standard Particle Model (SPM) are not able to describe all the phenomena in the Universe.

There are three possible ways to overcome these difficulties \citep{ref:star2}, \citep{ref:MDG6}. The first one is to add some new content in the Universe, like DM and DE).
The second one is to modify the theory of gravity. And the third one is some combination of the previous two.

Nowadays, the need for DM and DE is firmly established in CMB, large-scale structure formation, galaxy clusters, galaxy rotational curves, and
in accelerated expansion, high redshift type Ia supernovae, etc. (see, for example, \cite{ref:plank}), but their physical nature still remains a mystery.

One of the possibilities for modification of GR is through the f(R) theories. They generalize GR by replacing the scalar $R$ in the  Hilbert-Einstein action with some function $f(R)$. Unfortunately, the physical intuition cannot help find the explicit form of the function f(R) \citep{ref:buh}. 
At present we have not enough observational and experimental data to choose it.
As a result, a lot of such functions could be found in the literature. 
For example, \cite{ref:star1,ref:star2,ref:fR1,ref:fR2}. 
More extensive information about the $f(R)$ theories can be found in \cite{ref:book,ref:noj1,ref:noj2,ref:clif}.
For models of neutron stars in f(R) theories see, for example, \cite{ref:NS1, ref:NS2, ref:NS3, ref:NS4}.

The MDG as a proper generalization of the Einstein GR was first introduced by \citep{ref:ohan}.
His point was just to give some field-theoretical basis for the "fifth force" introduced by
\citep{ref:Fujii71} where the term "dilaton field" was used in this context for the first time.
Sometimes, this model is called also the "massive Branse-Dicke model with the parameter $\omega=0$" \citep{ref:alsing}.
Following \citep{ref:Fujii71}, the paper \citep{ref:ohan} used also the term dilaton for the Branse-Dicke scalar field $\Phi$.

Later on, the name MDG was introduced in \citep{ref:MDG} to distinguish the O'Hanlon model from other models with different kinds of dilaton fields used in different physical areas.
In \citep{ref:MDG,ref:MDG1} there were considered, also for the first time, astrophysical and cosmological applications of MDG, and the cosmological constant $\Lambda$ and cosmological units were introduced, to look for new areas of application of the simple O'Hanlon model.
After that the initial period of development of the MDG model, in \citep{ref:MDG1,ref:4D,ref:MDG2,ref:MDG3,ref:MDG4,ref:MDG5,ref:MDG6,ref:MDG7} its different
applications, as well as further justifications of the general theory were considered.

The MDG is a  simple extension of GR based on the following action of the gravi-dilaton sector:
\begin{equation}\label{eqn:MDGact}
S_{g,\Phi}=-\frac{c}{2k}\int \mathrm{d}^{4}x \sqrt{\vert g\vert}(\Phi R+2\Lambda U(\Phi)) ,
\end{equation}
where  $k=8\pi G/c^{2}$ is the Einstein constant, $G$ is the Newton gravitational constant, $\Lambda$ is the cosmological constant, 
and $ \Phi\in (0,\infty) $ is the gravitation-dilaton field. 
The values of $\Phi$ must be positive because a change of the sign would lead to a change of the sign of the gravitational factor 
$G/\Phi$, which would lead to antigravity. We rule out the possibility of antigravity, since it does not correspond to known real physical facts.
The value $\Phi =\infty$ must be excluded, because in this case the gravity is eliminated.
The value $\Phi =0$ is also unacceptable since it leads to an infinite gravitational factor, 
and the Cauchy problem is not well posed \citep{ref:EFarese01}.
\par
The Branse-Dicke gravitational dilaton field $\Phi$ is introduced in order to have a variable  gravitational factor $G(\Phi)=G/\Phi$ instead of gravitational constant $G$. 
The dilaton $\Phi$ does not enter in the action of the matter $S_{matter}$, because it has no interaction with ordinary matter of SPM.
Due to its specific physical meaning the dilaton $\Phi$ has unusual properties.

The function $U(\Phi)$ defines the cosmological potential and the whole extra dynamics of the model.
It is introduced in order to have a variable cosmological factor instead of the cosmological constant $\Lambda$.
The potential $U(\Phi)$ must be a singe valued function of the dilaton field due to astrophysical reasons.
If we set $\Phi=1$ and $U(\Phi)=1$, we are back into GR with the $\Lambda$ term.
\par
A special class of potentials are introduced in \citep{ref:MDG2}. They are called withholding potentials and they confine dynamically the values of the dilaton $\Phi$ in the physical domain.  In \citep{ref:MDG2} it is also shown that the MDG model is only locally equivalent to the $f(R)$ theories. 
The case of absence of such global equivalence leads to different physical consequences. 
Unfortunately, in the large existing literature one is not able to find functions $f(R)$ which are globally equivalent to the MDG model with the withholding potentials $U(\Phi)$.
There, only formal equivalence based on the Helmholtz approach in classical mechanics is sometimes discussed.
\par
In \citep{ref:MDG1}, MDG modifications of the classical GR effects in the solar system are considered: Nordvedt effect, Shapiro effect, perihelion shift, etc. 
In the weak field approximation, MDG is compatible with all known observational data if the mass of the dilaton $m_{\Phi}$ is large enough, i.e., if $m_\Phi > 10^{-3}\,\text{eV}$.
We also have an estimate from cosmology \citep{ref:star3} $m_{\Phi}\sim 10^{-6}M_{Plank}$.
The value of the dilaton mass $m_{\Phi}$ is the main open physical problem in MDG, as well as
in the locally equivalent to it $f(R)$ theories, and in other extended theories of gravity.

The field equation of MDG with matter fields can be found in \citep{ref:MDG,ref:MDG1,ref:MDG3,ref:MDG6,ref:MDG5}. They can be written in the following form, with $G=c=1$:
\begin{equation} \label{eqn:MDG1a}
\Phi G_{\alpha \beta}-\Lambda U(\Phi)g_{\alpha \beta}-\nabla_{\alpha}\nabla_{\beta}\Phi +g_{\alpha \beta}\Box \Phi=8\pi T_{\alpha \beta} ,
\end{equation}
\begin{equation}\label{eqn:MDG1b}
\Box \Phi+\Lambda V_{,\Phi}(\Phi)=\frac{8\pi}{3}T .
\end{equation}
\par
Here $T_{\alpha \beta}$ is the standard stress-energy tensor and $T$ is its trace. We use the standard notation for the Einstein tensor $G_{\alpha \beta}$. The dilatonic potential $V(\Phi)$ is introduced through its first derivative with respect of the dilaton, $V_{,\Phi}(\Phi)=\frac{2}{3}(\Phi U_{,\Phi}-2U)= \frac{2}{3}\Phi^{3}\frac{d}{d\Phi}(\Phi^{-2}U)$.
\par

In the papers \citep{ref:MDG3,ref:MDG4,ref:MDG5,ref:MDG6} there were considered models of static spherically symmetric neutron stars (NS) with different EOS of the matter:
ideal Fermi gas at zero temperature, polytropic EOS, and realistic EOS AMP1. Detailed derivation of the basic equations and the boundary conditions are given in \citep{ref:MDG6}.
Some general problems of singular and bifurcation manifolds for such stars were considered in \citep{ref:MDG7}.

The present paper deals with static NS with four other realistic EOS of matter: SLy, BSk19, BSk20, and BSk21.
It confirms and extends the basic results of the previous papers and completes the general picture of the MDG models of statically spherically symmetric NS.

 \section{Basic equations and boundary conditions for static spherically symmetric neutron stars}\label{s:basics}

With great precision, the static NS are spherically symmetric objects. 
So in the problem under consideration we can use the standard space-time interval, \citep{ref:Landau},
$$\mathrm{d}s^2=e^{\nu(r)}\mathrm{d}t^2-e^{\lambda(r)}\mathrm{d}r^2-r^2\mathrm{d}\theta^2 -r^2sin^{2}\theta \mathrm{d}\varphi^2,$$ 
where $r$ is the radial luminosity variable. The inner domain $r\in [0,r^{\star}]$, where $r^{\star}$ is the luminosity radius of the star, the structure is described by the following system of four first order differential equations, which represent the specific MDG generalization of the Tolman-Oppenheimer-Volkoff equations:
\begin{equation}\label{eqn:MDGM}
\frac{\mathrm{d}m}{\mathrm{d}r}=\frac{4\pi r^{2}\epsilon_{eff}}{\Phi} ,
\end{equation}
\begin{equation}\label{eqn:MDGP}
\frac{\mathrm{d}p}{\mathrm{d}r}=-\frac{(p+\epsilon)}{r\big( \Delta-2\pi r^3p_{\Phi}/ \Phi \big) }\Bigg( \frac{4\pi r^3 p_{eff}}{\Phi}+m\Bigg) ,
\end{equation}
\begin{equation}\label{eqn:MDGPhi}
\frac{\mathrm{d}\Phi}{\mathrm{d}r}=-\frac{4\pi r^2p_{\Phi}}{\Delta} ,
\end{equation}
\begin{equation}\label{eqn:MDGPPhi}
\frac{\mathrm{d}p_{\Phi}}{\mathrm{d}r}=-\frac{p_{\Phi}}{\Delta r}\Bigg( 3r-7m-\frac{2}{3}\Lambda r^3+\frac{4\pi r^3 \epsilon_{eff}}{\Phi}\Bigg) -\frac{2\epsilon_{\Phi}}{r} .
\end{equation}
\par
Here we have four unknown functions,  $m=m(r), p=p(r), \Phi=\Phi (r)$ and $p_{\Phi}=p_{\Phi}(r)$, the mass, the pressure, the dilaton and the dilaton pressure. 
The following notations are used in the system \eqref{eqn:MDGM}:
\begin{equation}	
	\begin{split}\label{ind}\nonumber
	\Delta &=r-2m-\frac{\Lambda r^3}{3} ,\\
	\epsilon_{eff}&=\epsilon+\epsilon_{\Lambda}+\epsilon_{\Phi}, \quad p_{eff}=p+p_{\Lambda}+p_{\Phi} ,\\
	\epsilon_{\Lambda}&=\frac{\Lambda}{8\pi}\big( U(\Phi)-\Phi\big), \quad p_{\Lambda}=-\frac{\Lambda}				{8\pi}\Bigg( U(\Phi)-\frac{\Phi}{3}\Bigg) ,\\
	\epsilon_{\Phi}&=p-\frac{1}{3}\epsilon +\frac{\Lambda}{8\pi}V\rq{}(\Phi)+\frac{p_{\Phi}\big( 					\frac{4\pi r^3}{\Phi}p_{eff}+m\big)}{2\big( \Delta-\frac{2\pi r^{3}p_{\Phi}}{\Phi}\big) } .
\end{split}		
\end{equation}

\begin{figure}[t]
\includegraphics[width=0.95\columnwidth, angle=0]{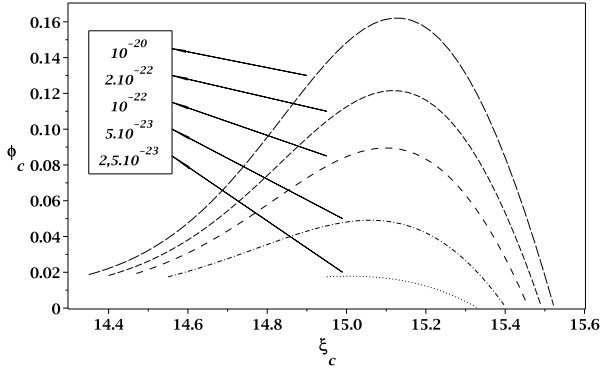}
\caption{Relation between the dilaton value and the density at the center for SLy EOS}
\label{fig:SLy phi-xi}
\end{figure}

In the above equation $\epsilon_{\Lambda}$ and $p_{\Lambda}$ are the cosmological energy density and cosmological pressure, $\epsilon_{\Phi}$ and $p_{\Phi}$ are the dilaton energy density and dilaton pressure. We combine cosmological, dilaton and matter energy density in a new variable $\epsilon_{eff}$. We do the same thing for the cosmological, dilaton and matter pressure in the variable $p_{eff}$.

\begin{figure}[t]
\begin{tabular}{ccc}
\includegraphics[width=0.95\columnwidth, angle=0]{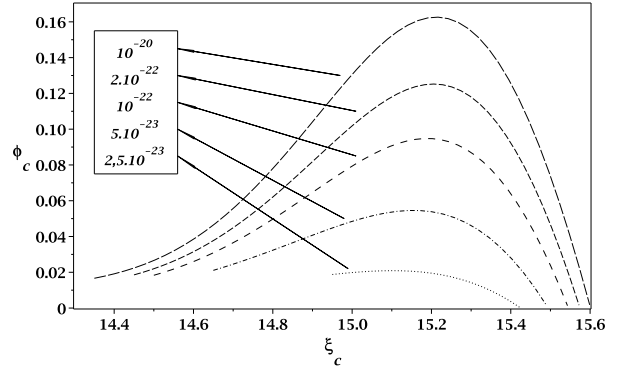} \\
\includegraphics[width=0.95\columnwidth, angle=0]{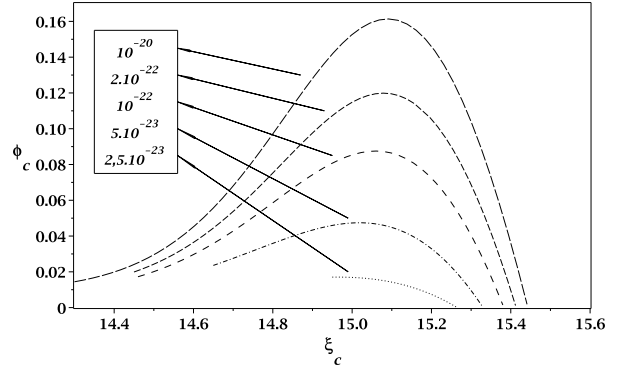} \\
\includegraphics[width=0.95\columnwidth, angle=0]{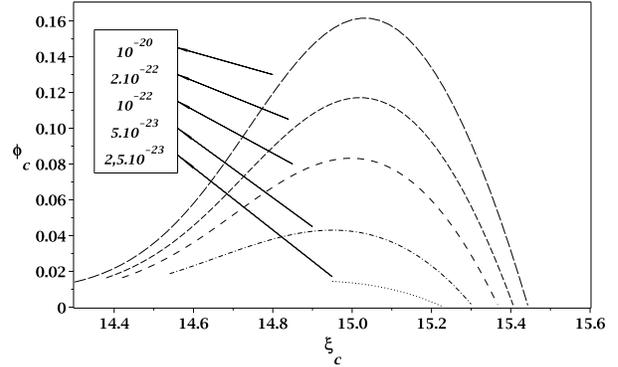}
\end{tabular}
\caption[Phi-xi relation]{Relation between the dilaton value and the density in the center. Top: BSk19. Middle: BSk20. Bottom: BSk21}
\label{fig:BSk phi-xi}
\end{figure}

In the present paper, we accept the standard assumption that the center of the spherically symmetric star is where the radial variable is zero, $r_{c}=0$. 
The obtained boundary conditions in the center of the star are \citep{ref:MDG6}:
\begin{eqnarray}\label{eqn:inbon}\nonumber
m(0)&=&0,\quad p(0)=p_c, \quad \Phi(0)=\Phi_c, \\
p_{\Phi}(0)&=&-\frac{2}{3}\Big( p_c-\frac{\epsilon_c}{3}\Big)-\frac{\Lambda}{12\pi}V_{,\Phi}(\Phi_c) .
\end{eqnarray}
On the edge of the star we impose the condition  $p^{\ast} =p(r^{\ast} ;p_c,\Phi_c)$ (and $\epsilon^{\ast}=0$). Then
\begin{eqnarray}\label{eqn:outbon}
m^{\ast}&=&m(r^{\ast};p_c,\Phi_c),\quad \Phi^{\ast}=\Phi(r^{\ast};p_c,\Phi_c),\\
 p^{\ast}_{\Phi}&=&p_{\Phi}(r^{\ast};p_c,\Phi_c).
\end{eqnarray}

\begin{table*}[t!]
\small
\caption{Configuration of the maximum allowable mass for non-rotating neutron star}
\label{tbl:MR}
\begin{tabular}{|c c c c c c c c c|}
\tableline
 & & & & & & & & \multicolumn{1}{c|}{} \\ [-1em]
           & \multicolumn{2}{c}{GR}     & \multicolumn{2}{c}{MDG $d=10^{-22}$} & \multicolumn{2}{c}{MDG $d=2.10^{-22}$} & \multicolumn{2}{c|}{MDG $d=10^{-20}$}  \\
EOS    & $ M[M_{\odot}] $ & R[km]  & $ M[M_{\odot}] $ & R[km]                          & $ M[M_{\odot}] $ & R[km]                                   &            $ M[M_{\odot}] $ & R[km]   \\   \tableline
SLy     &   2,05                    & 9,99    & 2,13                     & 10,44                            & 2,18                    & 10,68                           &             2,25                    & 11,01     \\
BSk19 &  1,86                    & 9,13     & 1,94                      & 9,59                             & 1,98                     & 9,80                           &             2,05                    & 10,02      \\
BSk20 &  2,15                    & 10,60    & 2,23                     & 10,74                           & 2,28                     & 10,92                           &             2,36                    & 11,17     \\
BSk21 &  2,28                    & 11,00    & 2,36                     & 11,59                           & 2,41                     & 11,80                           &             2,51                    & 12,03    \\
\tableline
\end{tabular}
\end{table*}

Outside of the star, where $p\equiv0$ and $\epsilon\equiv0$, we have a dilaton sphere or a dilasphere. The structure is determined by a shortened system \eqref{eqn:MDGM}$\div$\eqref{eqn:MDGPPhi}. Equation \eqref{eqn:MDGP} is omitted. In the exterior domain we use \eqref{eqn:outbon} as left boundary conditions. The right boundary conditions are defined by the cosmological horizon $r_U:$ $\Delta(r_U;p_c,\Phi_c)=0$ where the de Sitter vacuum is reached: $\Phi(r_U;p_c,\Phi_c)=1$.

\section{Model of NS with realistic equations of state SLy, BSk19, BSk20 and BSk21}\label{s:results}
\subsection{Equations of state}\label{EOS}

In the current research several realistic equations of state (EOS) are used to model neutron stars: SLy \citep{ref:SLy0,ref:SLy1}, BSk19, BSk20 and BSk21 \citep{ref:BSk0, ref:BSk1, ref:BSk2}. All EOS describe three qualitatively different regions of a neutron star. The outer crust, the inner crust, and the core of the star, which are separated by phase transition points. All of the considerate EOS are compatible with the latest observational data, for the maximum mass of neutron star \footnote{EOS BSk19 gives maximum mass $M_{max}=1.86M_{\odot}$ in GR, which is below the maximum observed mass of neutron stars, but in MDG, the obtained masses for BSk19 are compatible with the observational data}. \citep{ref:dem, ref:ant}.
\par
 For the purpose of numerical simulations in the current paper we have used their analytical representations \citep{ref:SLyan} and \citep{ref:BSkan}. Considering the different character of the EOS in the different regions that they describe, the resulting fit is quite complicated. They parametrized the pressure as a function of the density, and more precisely their logarithmic values $\xi=\log (\rho / g~cm^{-3})$ and $\zeta=\log (P / dyn~cm^{-2})$. The typical error of the fit is $1-2\%$, and the maximum is $3.7\%$.

\subsection{Results}
Before we integrate the MDG equations describing neutron stars for given EOS, we have to clarify the explicit form of the cosmological potential $U(\Phi)$. The simplest withholding dilaton potential can be written in the following form \citep{ref:MDG1,ref:4D,ref:MDG2}:
\begin{equation}\label{eqn:pot}
U(\Phi)=\Phi^{2}+\frac{3}{16d^{2}}(\Phi-1/ \Phi)^{2} ,
\end{equation}
where the dimensionless Compton length $d=\lambda_{\Phi}\sqrt{\Lambda}$ is used; $\Lambda$ is the cosmological constant and $\lambda_{\Phi}$ is the dilaton Compton length.
\par

In order to have a successful computation, with high precision the dilaton field $\Phi$ is replaced by a new variable $\phi$ \citep{ref:MDG6}
\begin{equation}\label{phi}
\phi=\ln(1+\ln\Phi)\Leftrightarrow \Phi=\exp(\exp(\phi)-1) .
\end{equation}
The new double logarithmic scale stretches the physical domain of the scalar field and greatly expands the possibilities for numerical calculations.

\par

The first step in solving this physical problem is to obtain the initial conditions. The dependence between the dilaton field and the density, at the center,  is unique for the different EOS, and for different values of the dilaton Compton length. Figure \ref{fig:SLy phi-xi} shows the results for SLy EOS and Fig. \ref{fig:BSk phi-xi} for the BSk19, BSk20 and BSk21 EOS. For different realistic EOS (see also Fig.2 in \citep{ref:MDG6}) the qualitative behavior of the functions $\Phi_{c}(\xi_{c};\lambda_{\Phi})$ is similar, but the numerical values are different, as one can see on Fig.\ref{fig:SLy phi-xi}. For values of $d>10^{-20}$ the results are very close to ones $d=10^{-20}$. Further increment of the parameter $d$ does not lead to substantially new results. A feature of the MDG model is the shortening of the physical domain, in all EOS, as the Compton length decreases. This is probably due to a bifurcation in the equations \eqref{eqn:MDGM}$\div$\eqref{eqn:MDGPPhi}. This problem is partially discussed in \citep{ref:MDG6,ref:MDG7}.

\begin{figure*}[t!]
\begin{tabular}{lcr}
\includegraphics[width=0.62\columnwidth, angle=0]{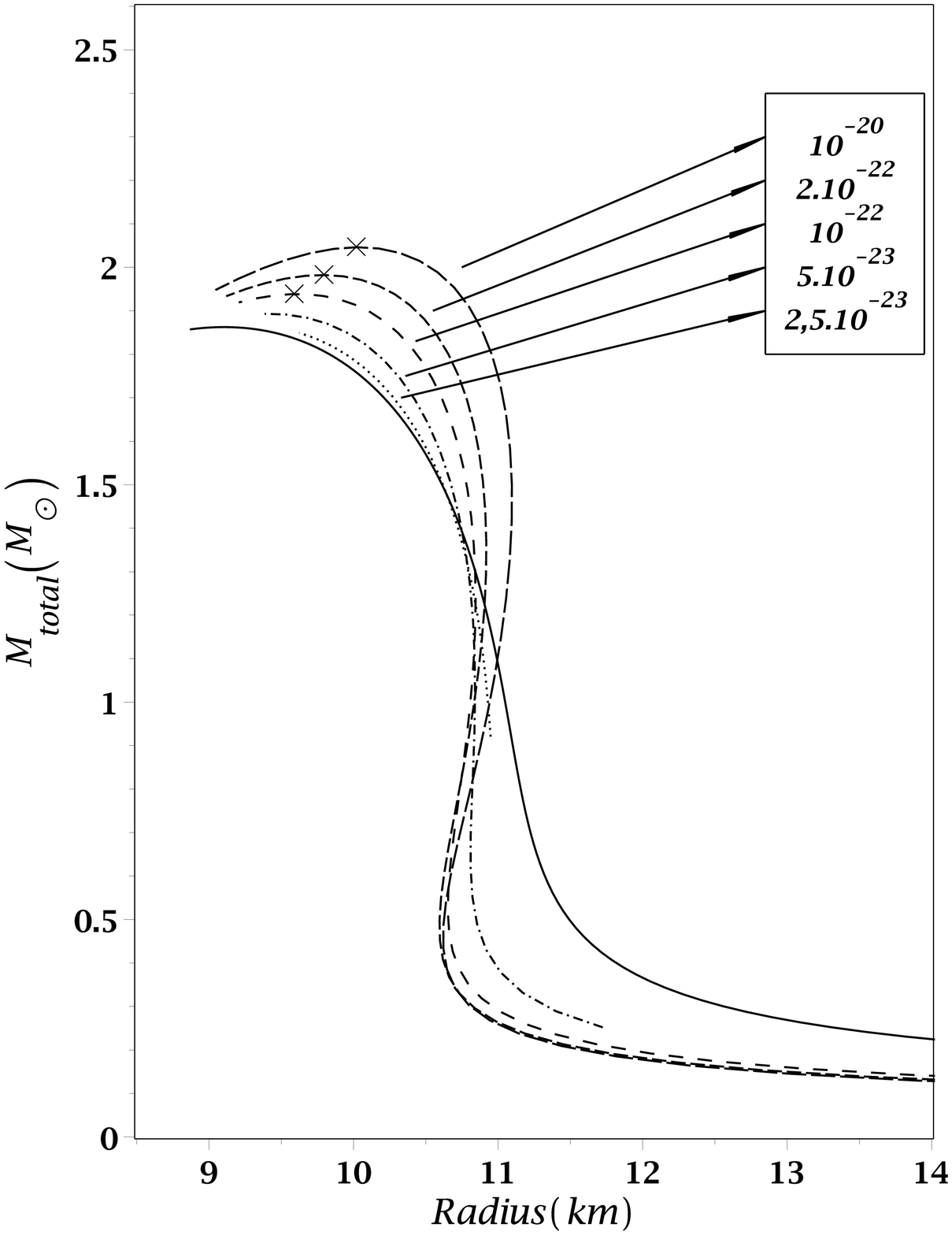} &
\includegraphics[width=0.62\columnwidth, angle=0]{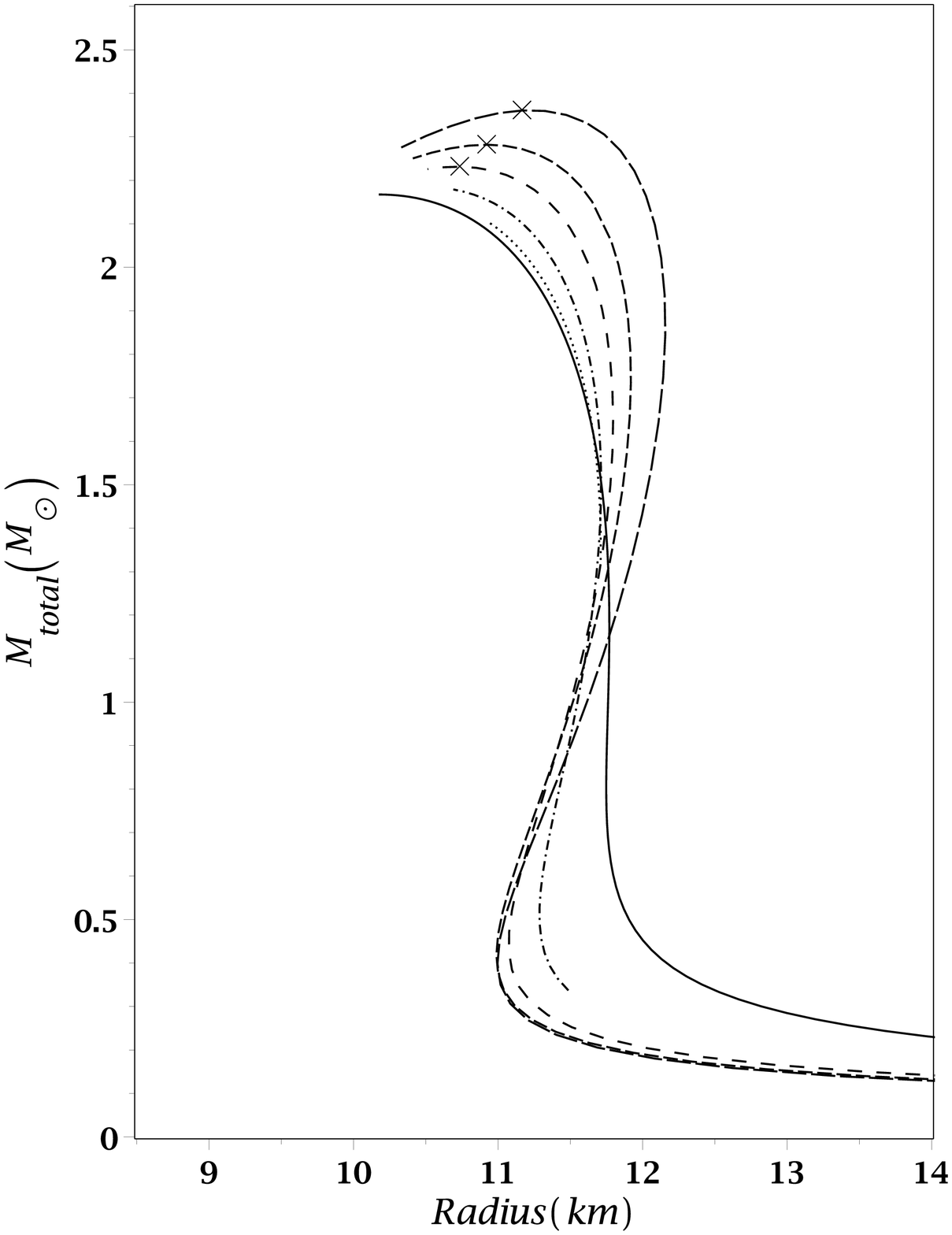} &
\includegraphics[width=0.62\columnwidth, angle=0]{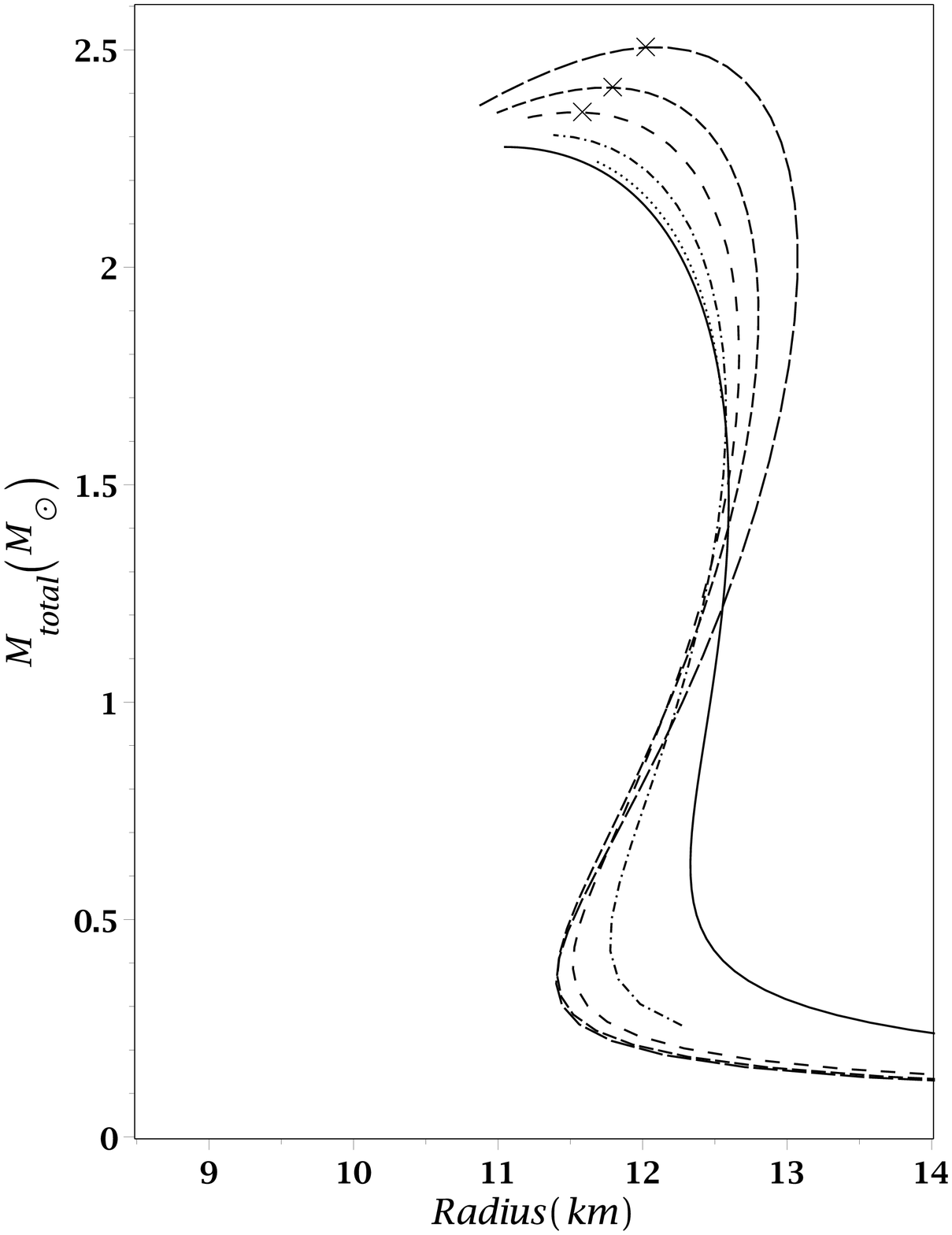}
\end{tabular}
\caption{Mass-radius relations for BSk19 (left), BSk20 (middle) and BSk21 (right). The crosses mark the threshold, beyond which the neutron stars are unstable. The solid line is the mass-radius relation from GR. The different line styles represent the same dimensionless parameter $d$ on all three panels}
\label{fig:BSkMR}
\end{figure*}

\begin{figure}[t!]
\includegraphics[width=0.95\columnwidth, angle=0]{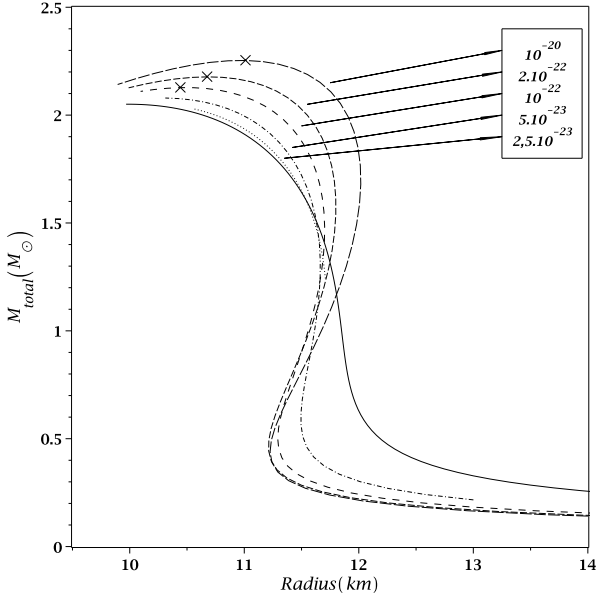}
\caption{Mass-radius relation for SLy EOS.  The crosses mark the threshold, beyond which the neutron stars are unstable. The solid line is the mass-radius relation from GR}
\label{fig:SLyMR}
\end{figure}

\begin{figure}[t!]
\begin{tabular}{cc}
\includegraphics[width=0.95\columnwidth, angle=0]{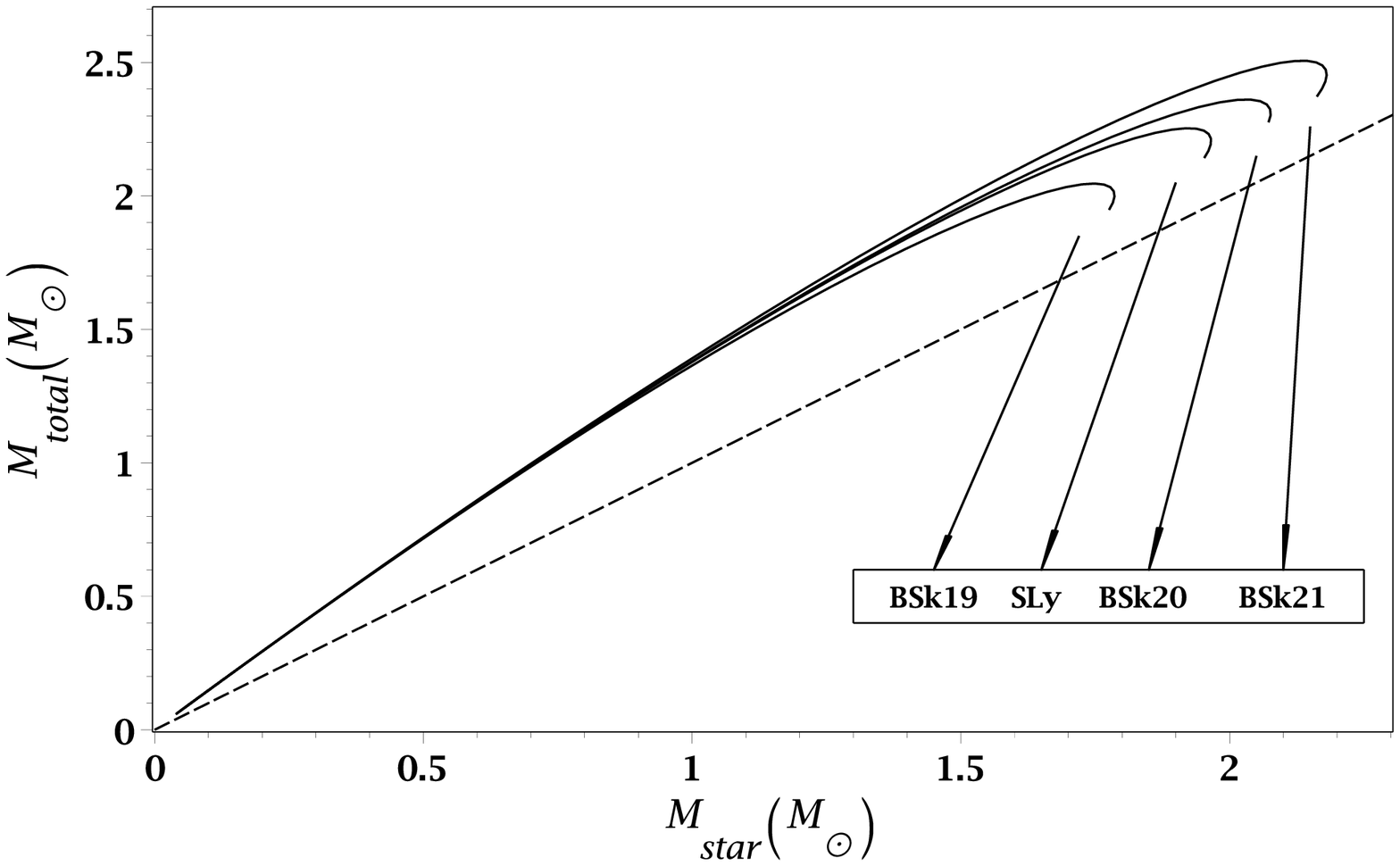} \\
\includegraphics[width=0.95\columnwidth, angle=0]{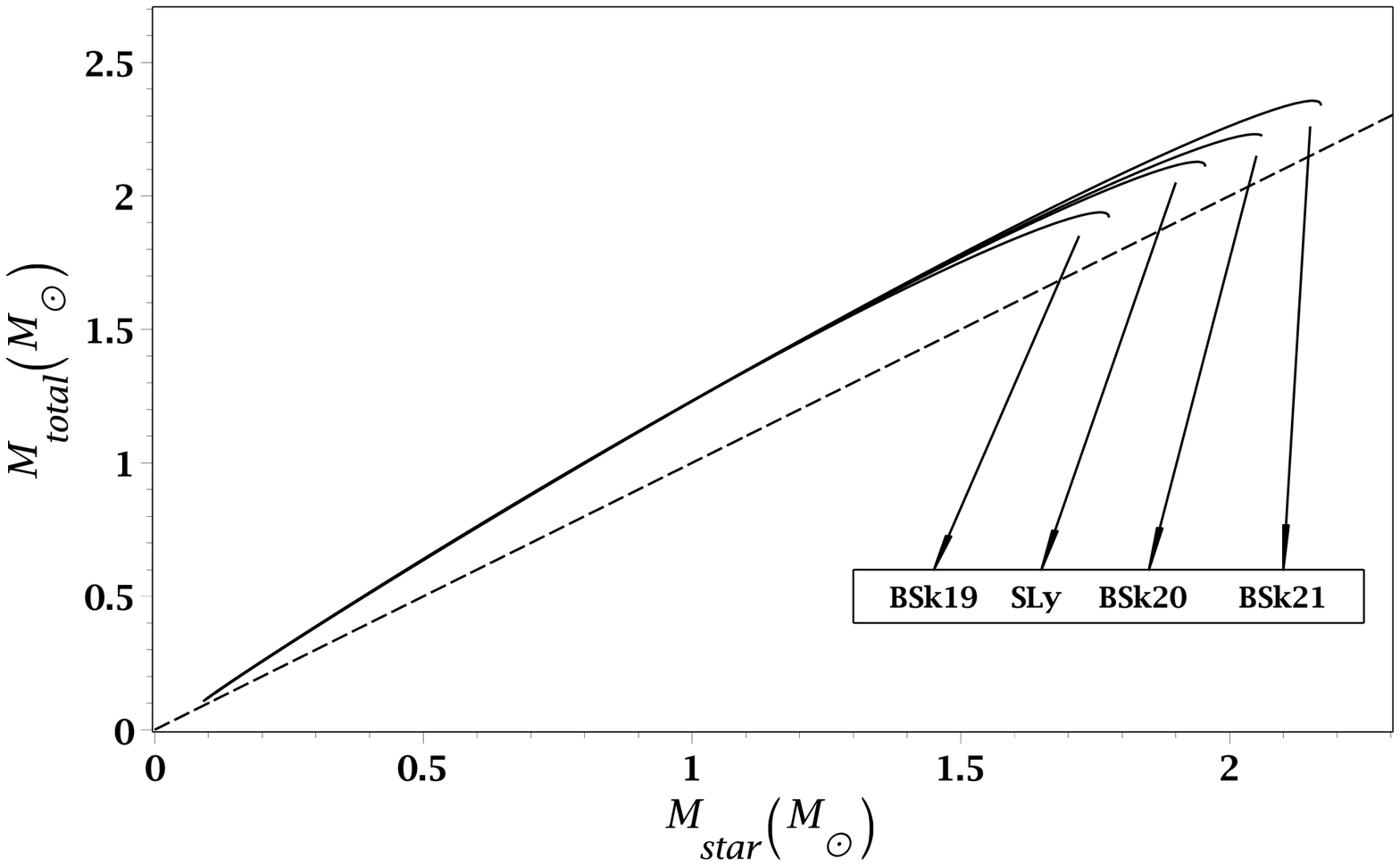}
\end{tabular}
\caption{Total mass of the star in MDG (mater and dilasphere) versus neutron star mass in MDG without the dilasphere, for $d=10^{-20}$ (upper panel) and $d=10^{-22}$ (lower panel). The dashed line is $M_{total}=M_{star}$ and it is for comparison}
\label{fig:MM}
\end{figure}

Using the obtained initial conditions for the different EOS and for different Compton lengths, we integrate the system \eqref{eqn:MDGM}$\div$\eqref{eqn:MDGPPhi}. Figure \ref{fig:BSkMR} shows the mass-radius relations for non-rotating neutron stars in the MDG model for EOS BSk19, BSk20, BSk21. The results for SLy EOS are shown on Fig. \ref{fig:SLyMR}. For all EOS the MDG model gives higher maximum masses than the corresponding GR solution. The results are summarized in Table. \ref{tbl:MR}. For the smallest studied value of $d$, the results are very close to those from GR. For the highest studied $d$ the maximum mass is $ \sim 10\%$ higher. The bigger maximum mass is due to the new feature in MDG neutron stars - the dilasphere. On Fig.\ref{fig:MM}, the difference between the total mass of the star (matter and dilasphere) and the mass of the neutron star without the dilasphere is presented. It can be seen that the impact of the dilashpere on the full mass of the object is significant. The mass of the dilasphere varies in accordance with the mass of the star. It is between $15\div 30\%$ of the mass of the entire object. The heaviest dilasphere is $M_{dil}\approx 0.5M_{\odot}$.

\par

\begin{figure}[t!]
\begin{tabular}{ccc}
\includegraphics[width=0.95\columnwidth, angle=0]{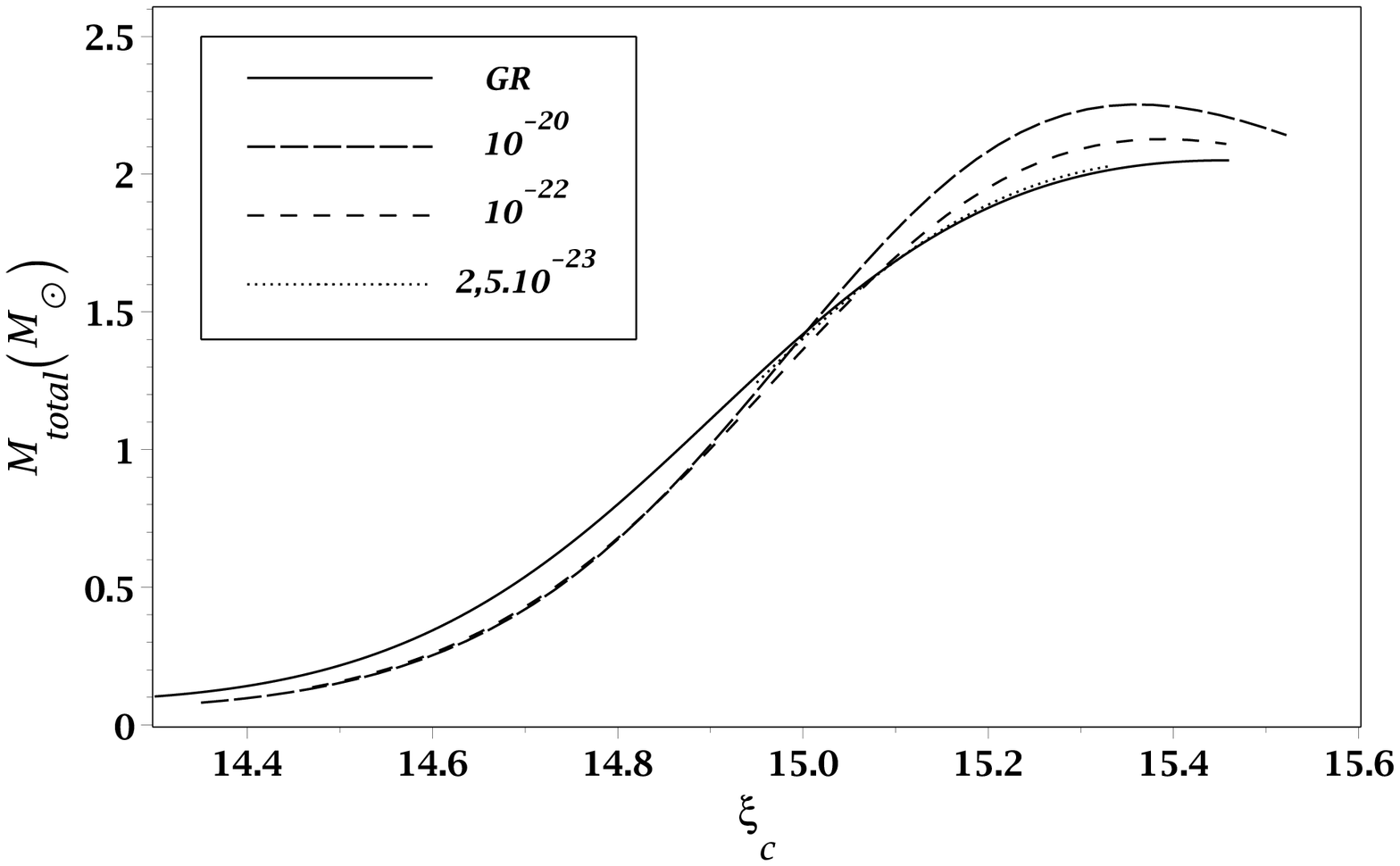} \\
\includegraphics[width=0.95\columnwidth, angle=0]{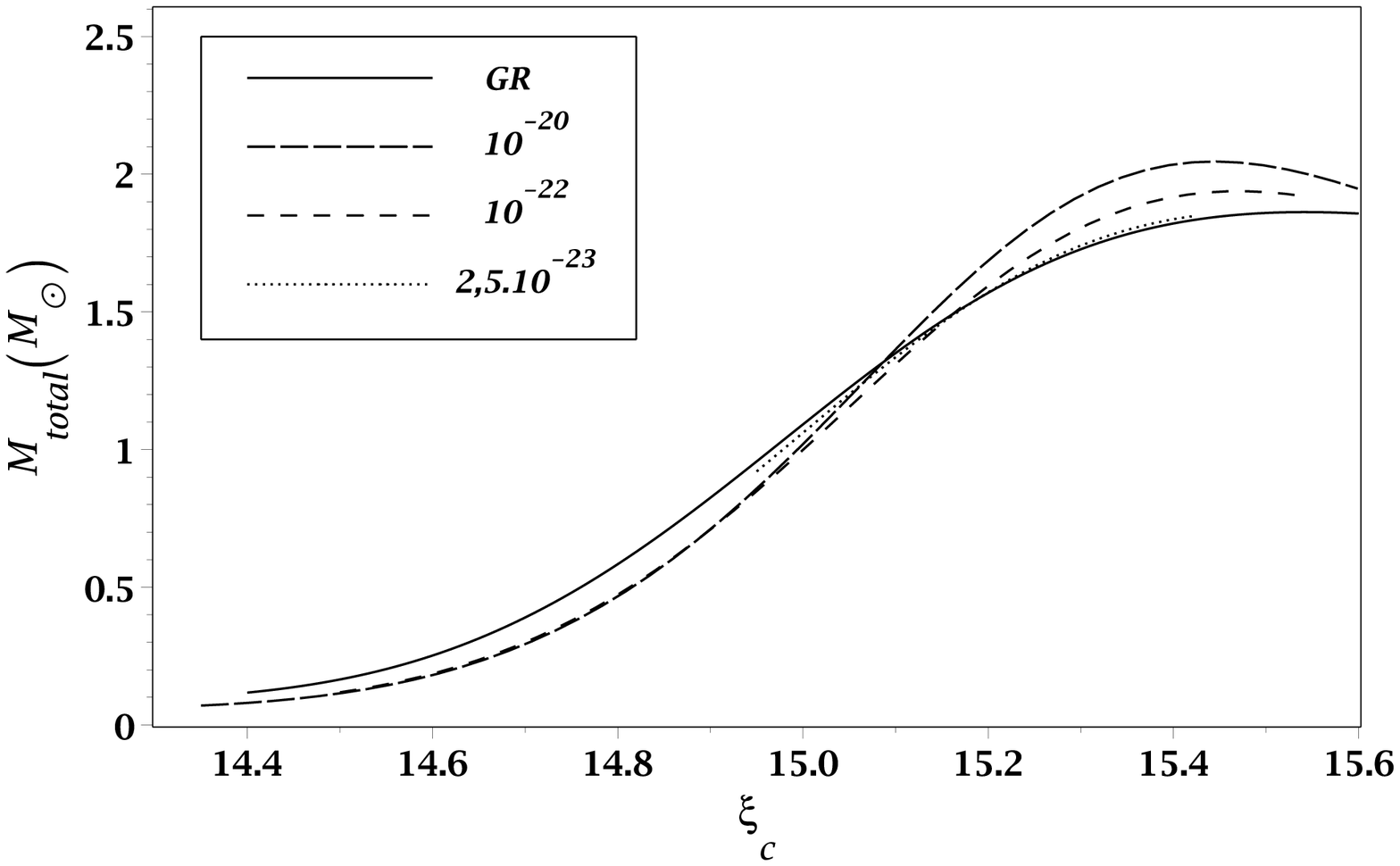} \\
\includegraphics[width=0.95\columnwidth, angle=0]{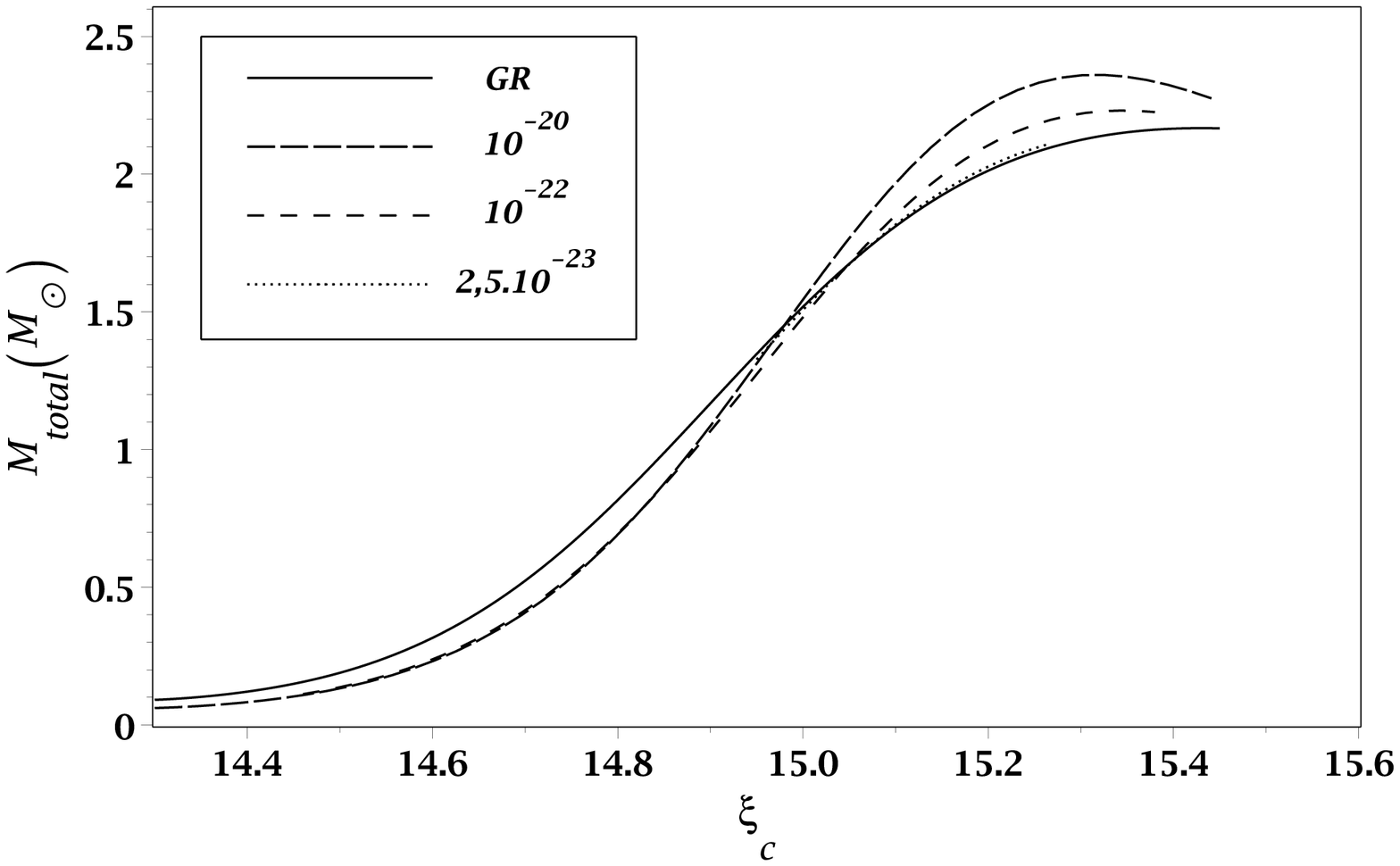} \\
\includegraphics[width=0.95\columnwidth, angle=0]{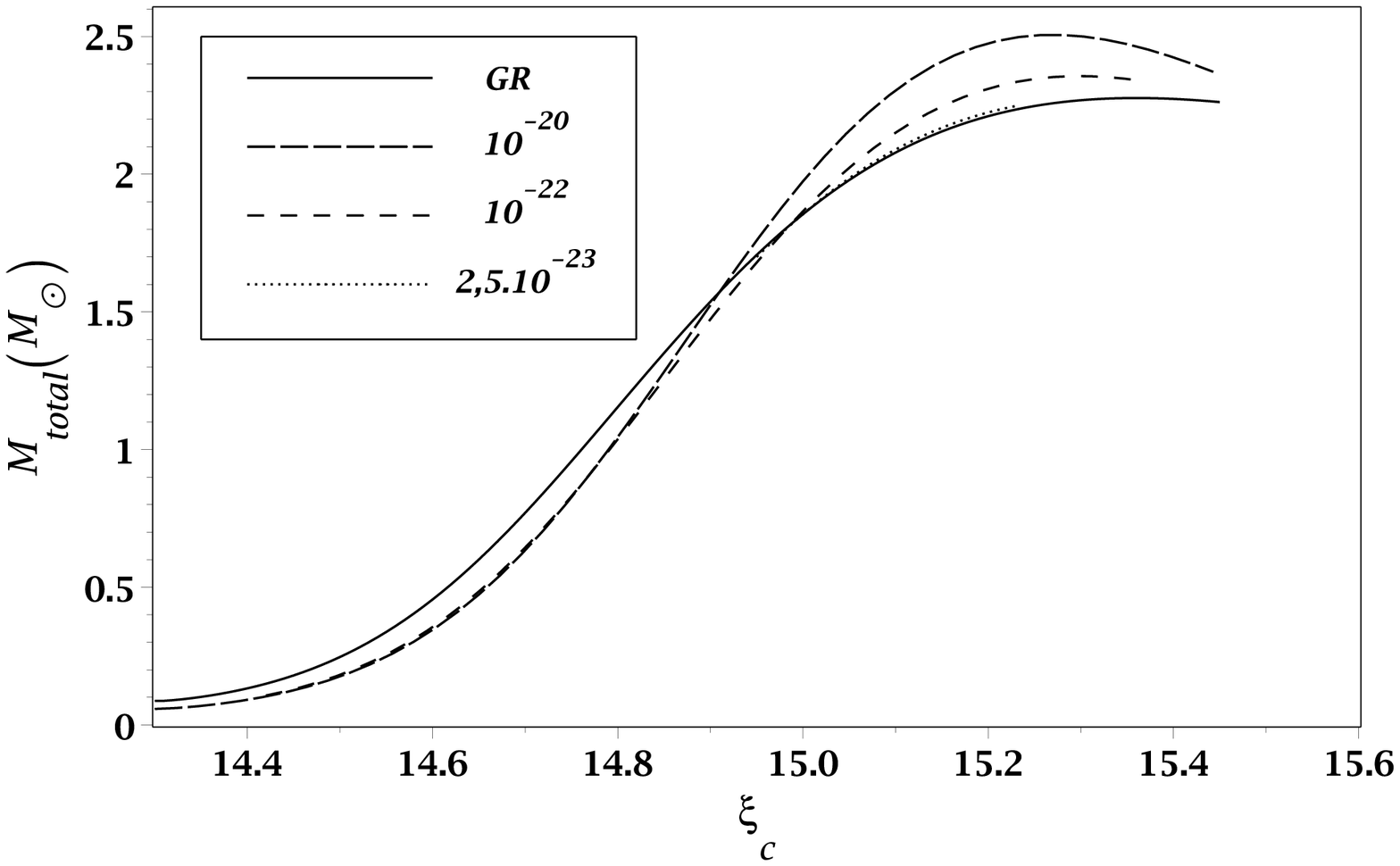}
\end{tabular}
\caption[M-xi relation]{Relation between the total mass of the star and the density in the center. From top to bottom the panels are for SLy, BSk19, BSk20, BSk21 EOS}
\label{fig:EOSMxi}
\end{figure}

Figure \ref{fig:EOSMxi} represents the relation between the total mass of the star $M_{total}$ and the density at the center of the star $\xi_{c}=\log \rho_{c}$. For lower central densities the results obtained from GR give bigger masses, but for high enough central density MDG, for all values of the dilaton Compton length, gives bigger masses, including the maximum one. This statement is valid for all the realistic EOS that we have used.

\begin{figure}[t!]
\begin{tabular}{cc}
\includegraphics[width=0.95\columnwidth, angle=0]{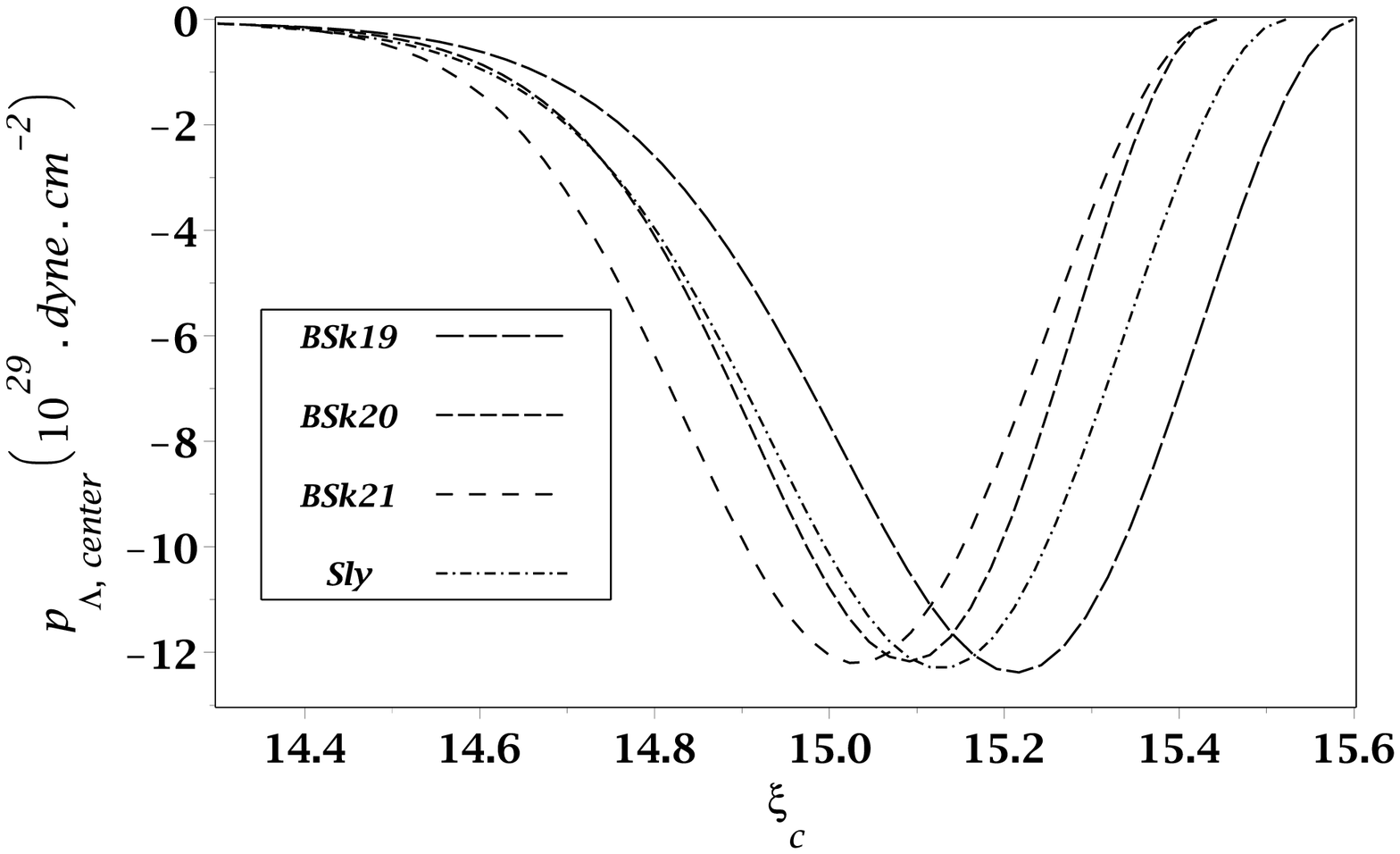} \\
\includegraphics[width=0.95\columnwidth, angle=0]{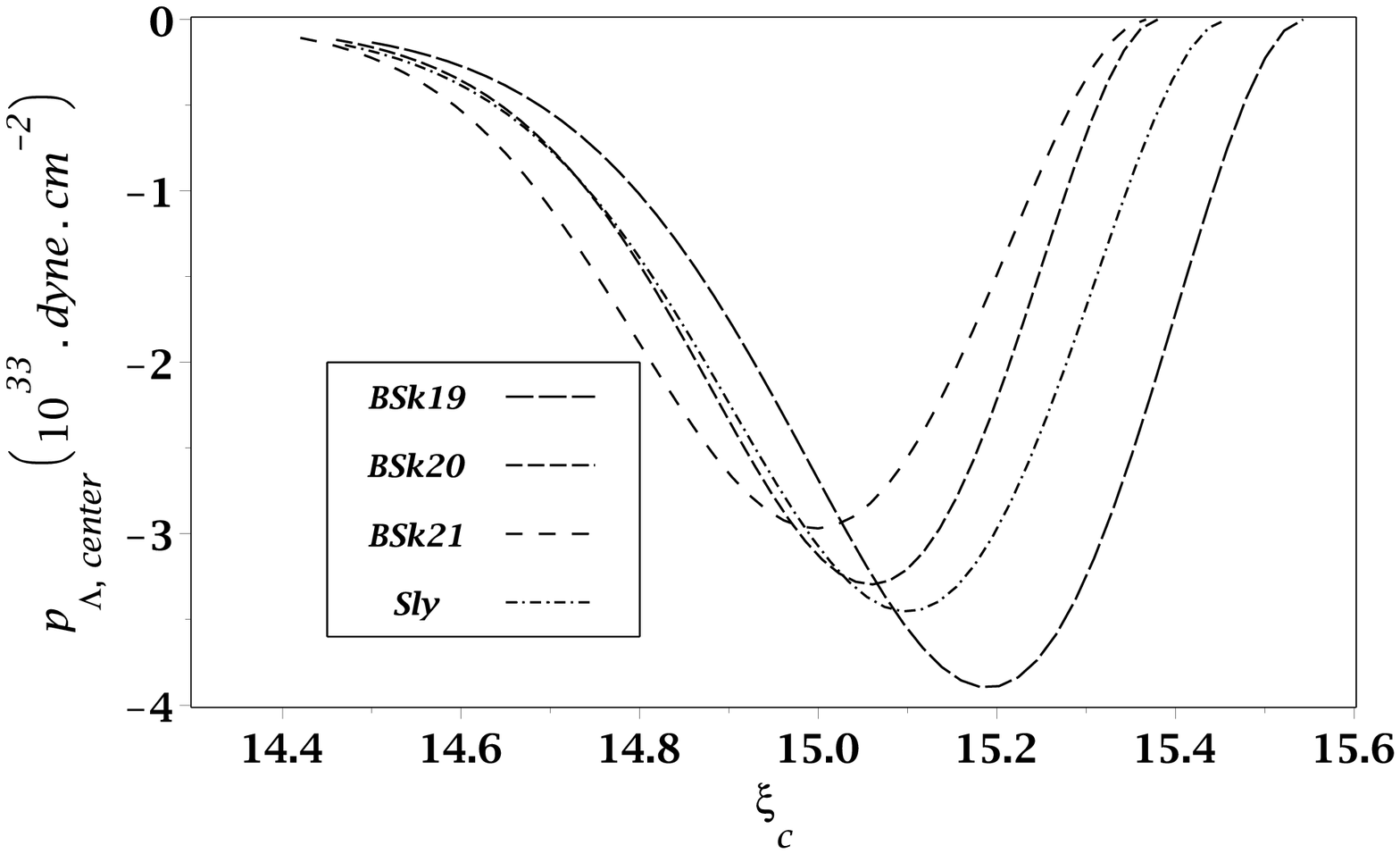}
\end{tabular}
\caption{Cosmological pressure $p_{\Lambda}$ in the center of the star as function of the central density, for $d=10^{-20}$ (upper panel) and $d=10^{-22}$ (lower panel)}
\label{fig:plc}
\end{figure}

\begin{figure}[t!]
\begin{tabular}{cc}
\includegraphics[width=0.95\columnwidth, angle=0]{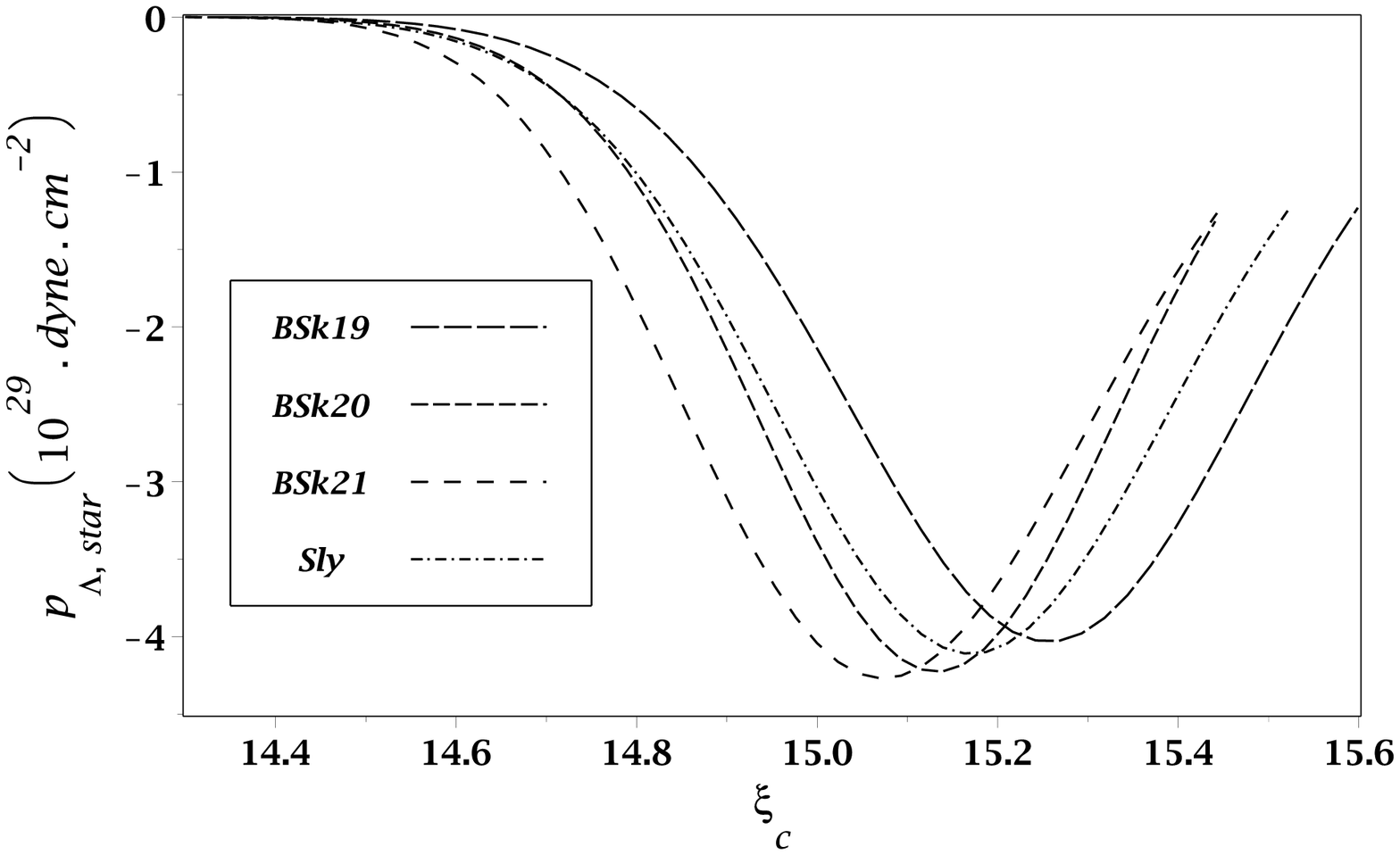} \\
\includegraphics[width=0.95\columnwidth, angle=0]{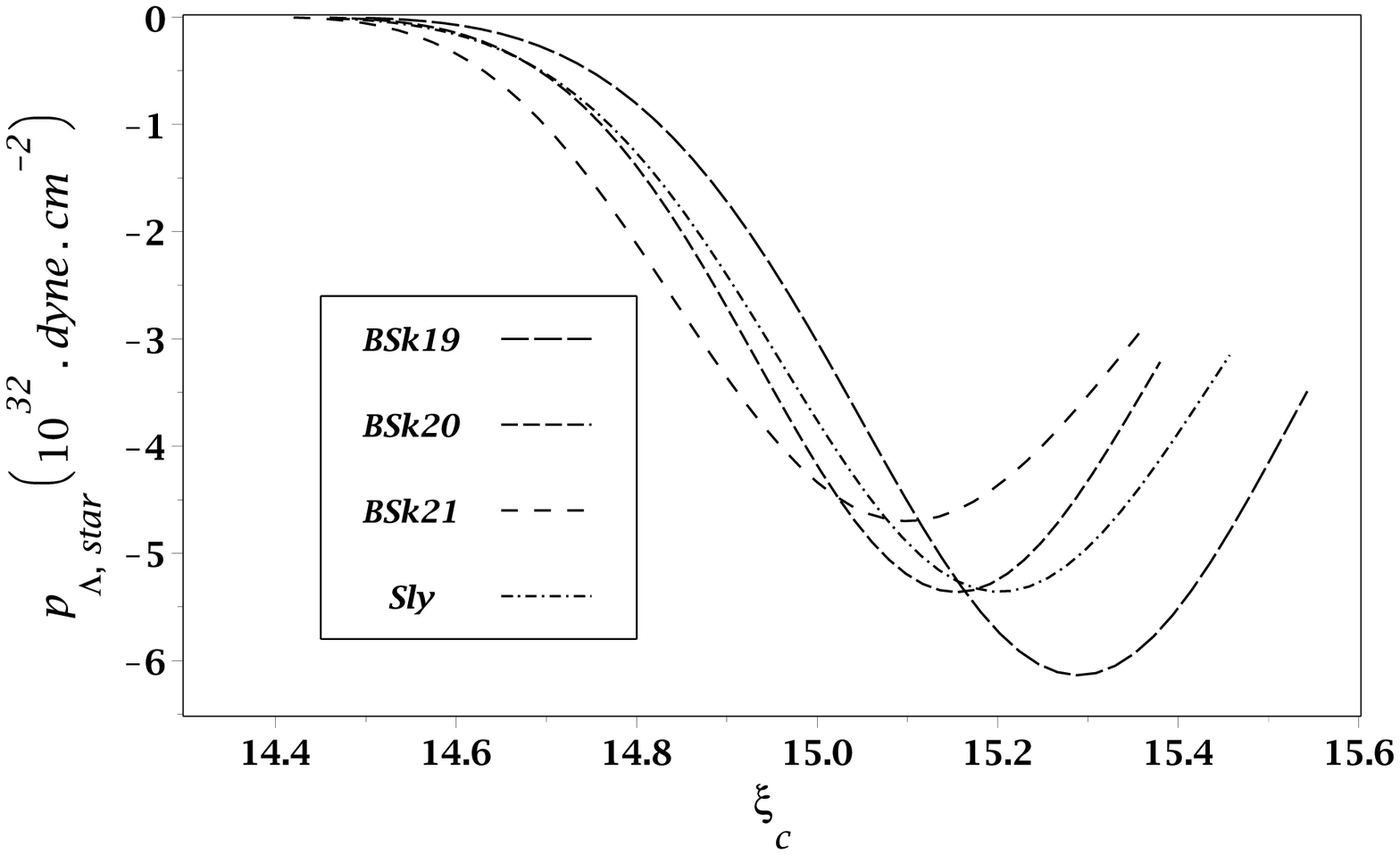}
\end{tabular}
\caption{Cosmological pressure $p_{\Lambda}$ on the edge of the star as a function of the central density, for $d=10^{-20}$ (upper panel) and $d=10^{-22}$ (lower panel)}
\label{fig:pls}
\end{figure}

The bigger mass is not solely due to the additional mass, which the dilaton brings in the system. In MDG the effective pressure of the star is the sum of three different components $ p_{eff}=p+p_{\Lambda}+p_{\Phi}$, the matter pressure $p$, the cosmological pressure $p_{\Lambda}$, and the dilaton pressure $p_{\Phi}$. The cosmological pressure $p_{\Lambda}$ is negative for all the EOS that we have used here, for all initial conditions. On Fig.\ref{fig:plc}, $p_{\Lambda}$ in the center of the star is plotted as a function of the central density, and on Fig.\ref{fig:pls}, $p_{\Lambda}$ on the edge of the star is plotted as a function of the central density. The dilaton pressure increases towards the edge but still remains negative with a significant value on the edge. This effect is similar to what we can expect from dark energy.
\par

\begin{figure}[t!]
\begin{tabular}{cc}
\includegraphics[width=0.95\columnwidth, angle=0]{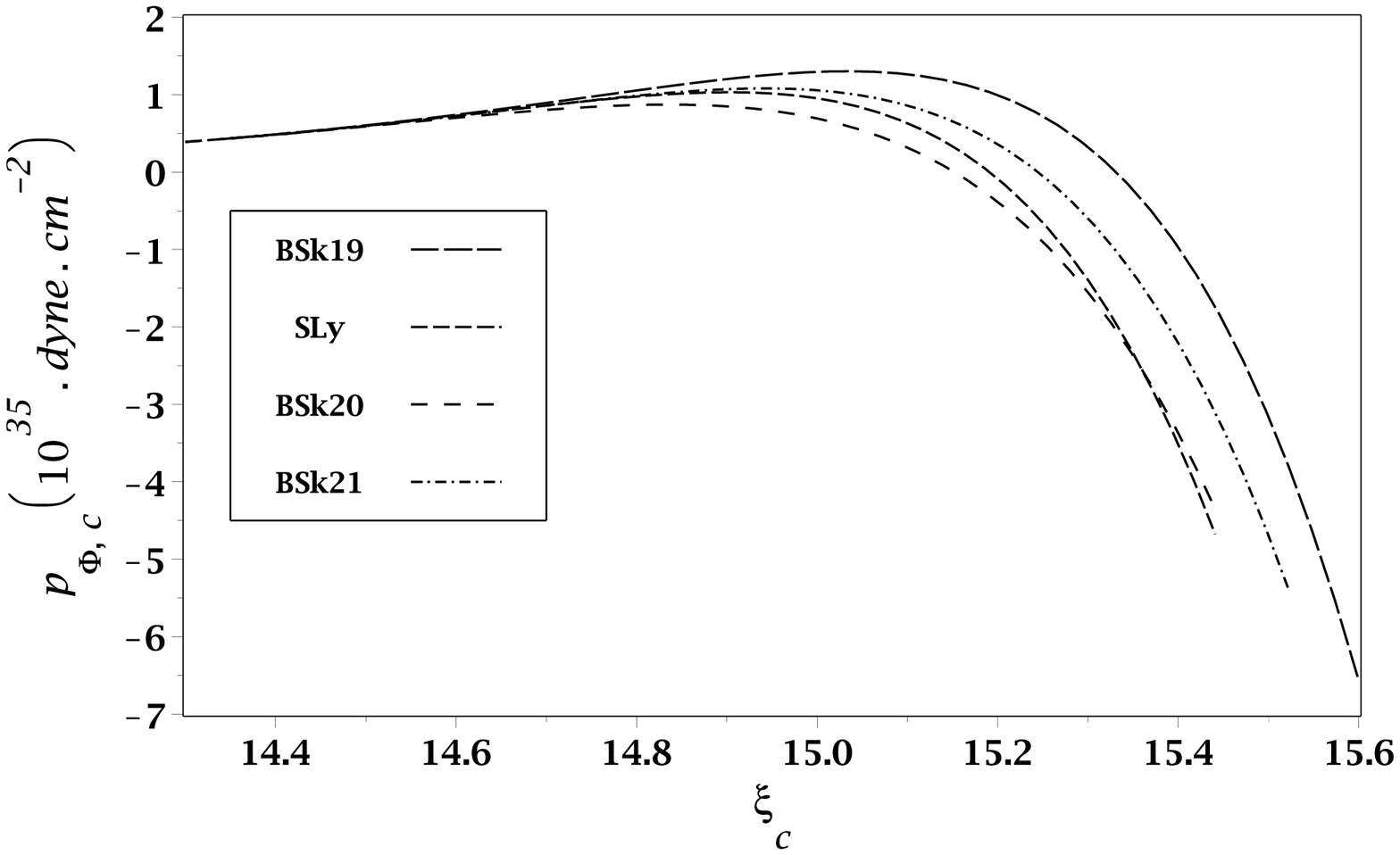} \\
\includegraphics[width=0.95\columnwidth, angle=0]{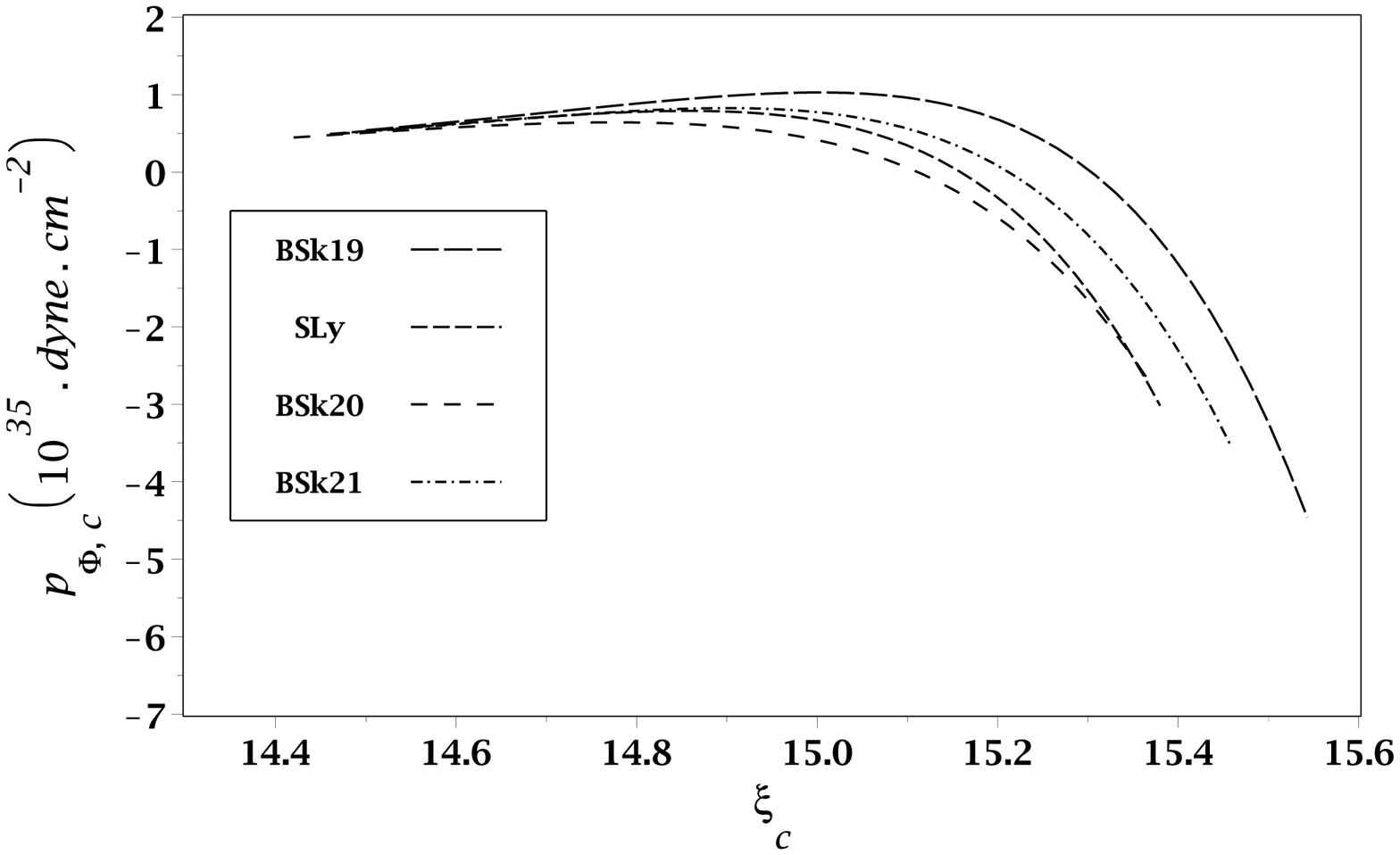}
\end{tabular}
\caption{Dilaton pressure $p_{\Phi}$ in the center of the star as a function of the central density, for $d=10^{-20}$ (upper panel) and $d=10^{-22}$ (lower panel)}
\label{fig:ppc}
\end{figure}

\begin{figure}[t!]
\begin{tabular}{cc}
\includegraphics[width=0.95\columnwidth, angle=0]{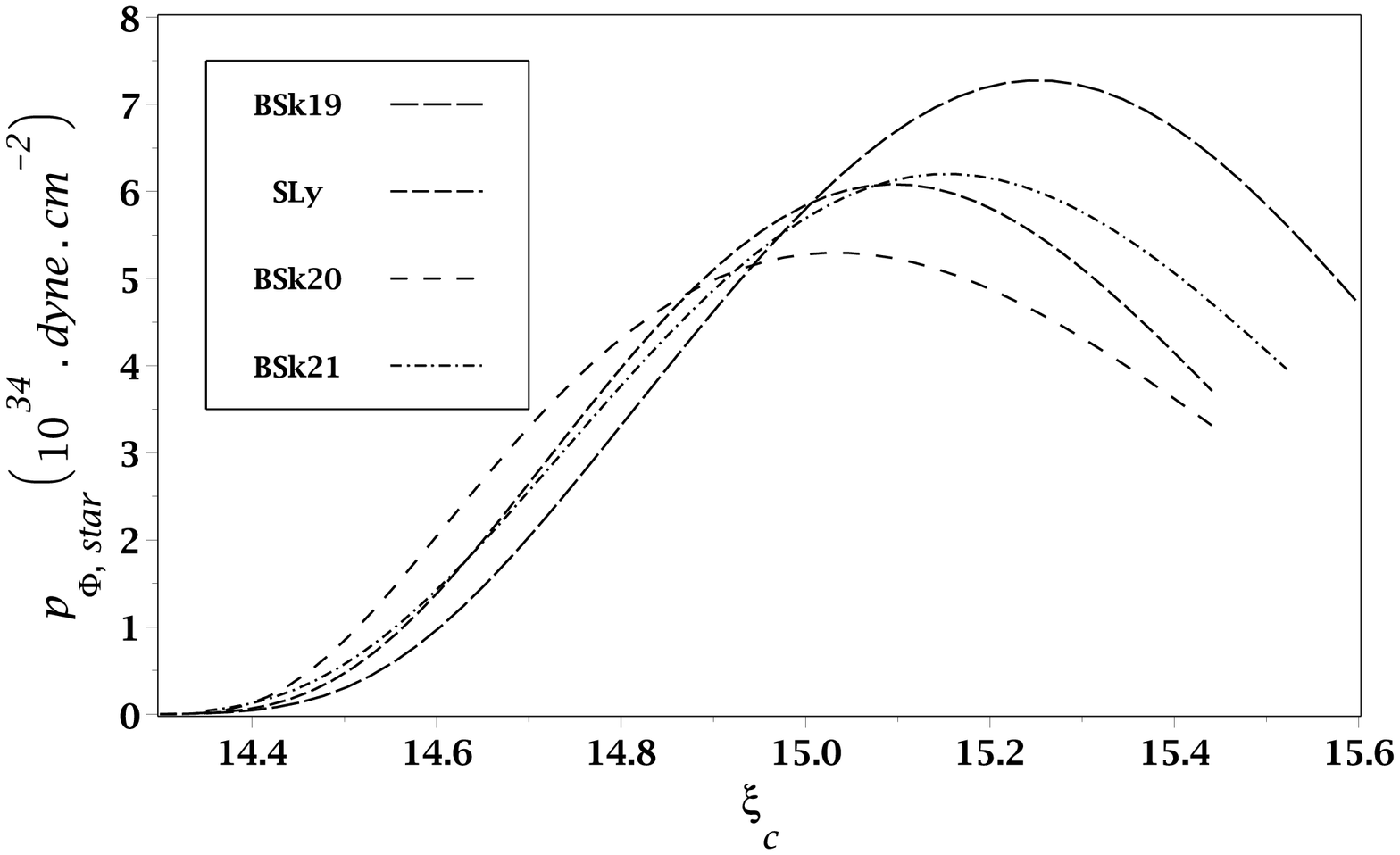} \\
\includegraphics[width=0.95\columnwidth, angle=0]{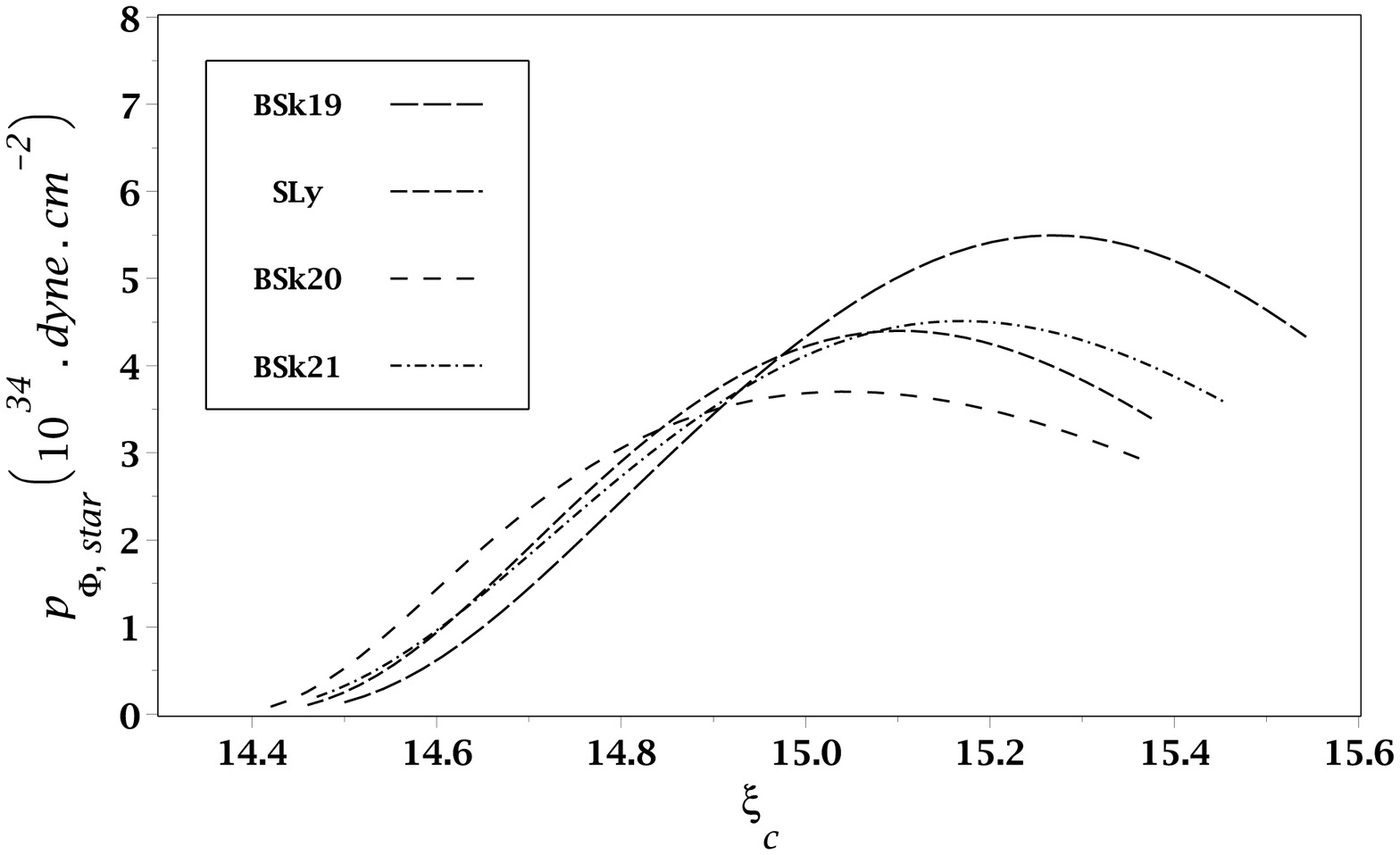}
\end{tabular}
\caption{Dilaton pressure $p_{\Phi}$ on the edge of the star as a function of the central density, for $d=10^{-20}$ (upper panel) and $d=10^{-22}$ (lower panel)}
\label{fig:pps}
\end{figure}

The dilaton pressure $p_{\Phi}$ at the center of the star can have positive and negative values depending on the initial conditions, Fig.\ref{fig:ppc}. It is positive for low central densities and becomes negative for the highest physically meaningful values of the central density, for the different EOS. This behavior of $p_{\Phi}$ is the same for all EOS we have used in the present paper. On the edge of the star the dilaton pressure is always positive $p_{\Phi}>0$, for all central densities, Fig.\ref{fig:pps}. This is another proof of the concept that a dilasphere exists around neutron stars in the model of MDG.  The dilaton does not interact with ordinary matter but adds additional mass to the neutron star, and has pressure $p_{\Phi}\neq 0$ inside the star and outside the star, to some extent. These effects can be interpreted as dark matter.
\par
Additional information on the behavior of the different types of pressure in a single neutron star and its dilasphere can be found in \citep{ref:MDG3, ref:MDG4, ref:MDG5, ref:MDG6}. 
\par

For the cases of EOS considered here, the cosmological energy density $\epsilon_{\Lambda}$, contrary to $p_{\Lambda}$, is always positive. In the center, Fig.\ref{fig:elc} and on the edge of the star, Fig.\ref{fig:els}. The dilaton energy density $\epsilon_{\Phi}$ in the center of the star is negative for small central densities but it becomes positive for larger ones, exactly the opposite behavior compared to $p_{\Phi}$, Fig.\ref{fig:epc}. On the edge of the star $\epsilon_{\Phi}$ is positive for all central densities Fig.\ref{fig:eps}. 
\par
The contribution of the cosmological and dilaton variables plays a very important role in the star physics in MDG. Their contribution is not negligible and it makes the difference between a neutron star in GR and a neutron star in MDG. The behavior of $p_{\Lambda}, \epsilon_{\Lambda}$ and $p_{\Phi}, \epsilon_{\Phi}$ helps us to interpret them as possible candidates for DE and DM.
\par
\begin{figure}[t!]
\begin{tabular}{cc}
\includegraphics[width=0.95\columnwidth, angle=0]{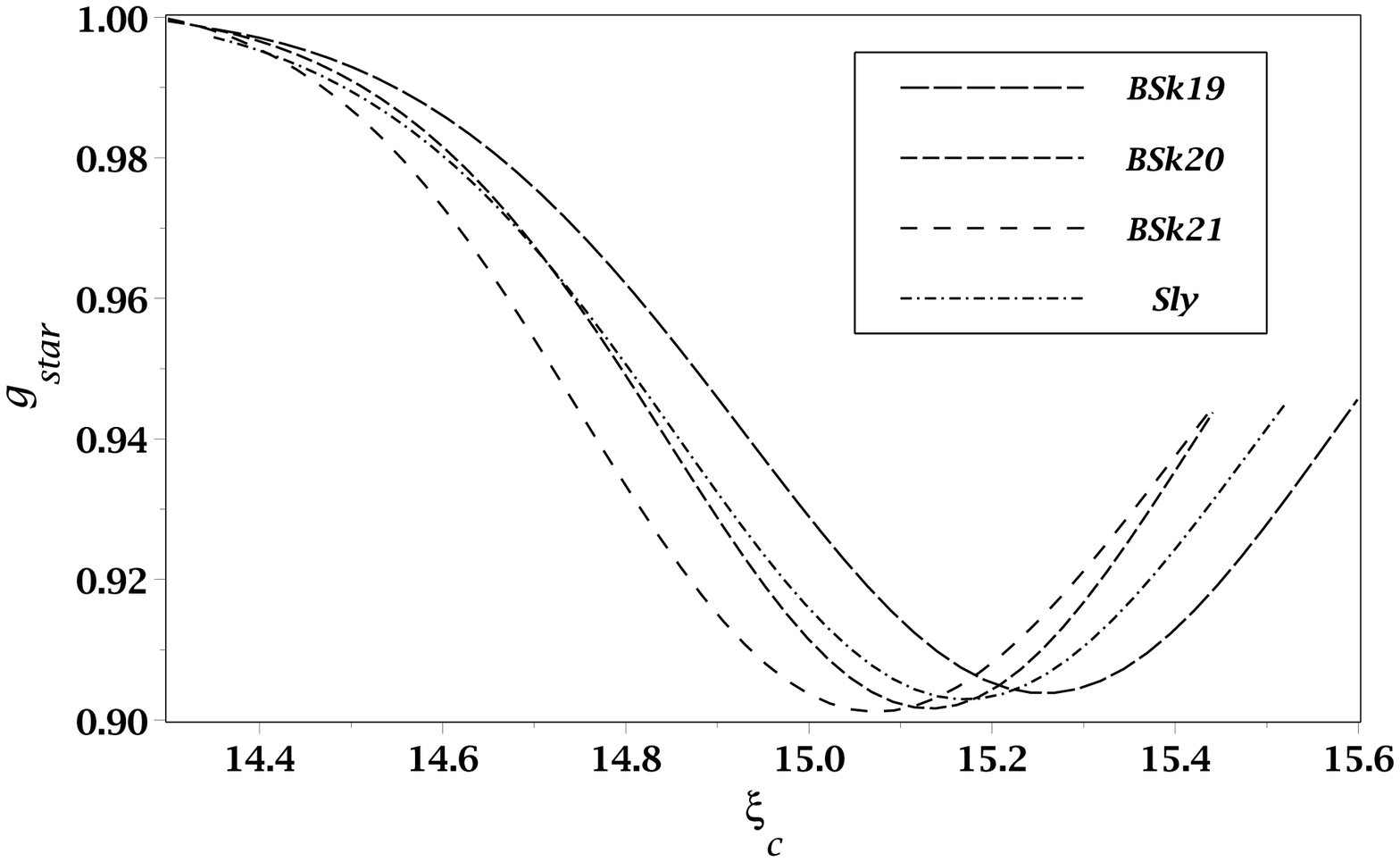} \\
\includegraphics[width=0.95\columnwidth, angle=0]{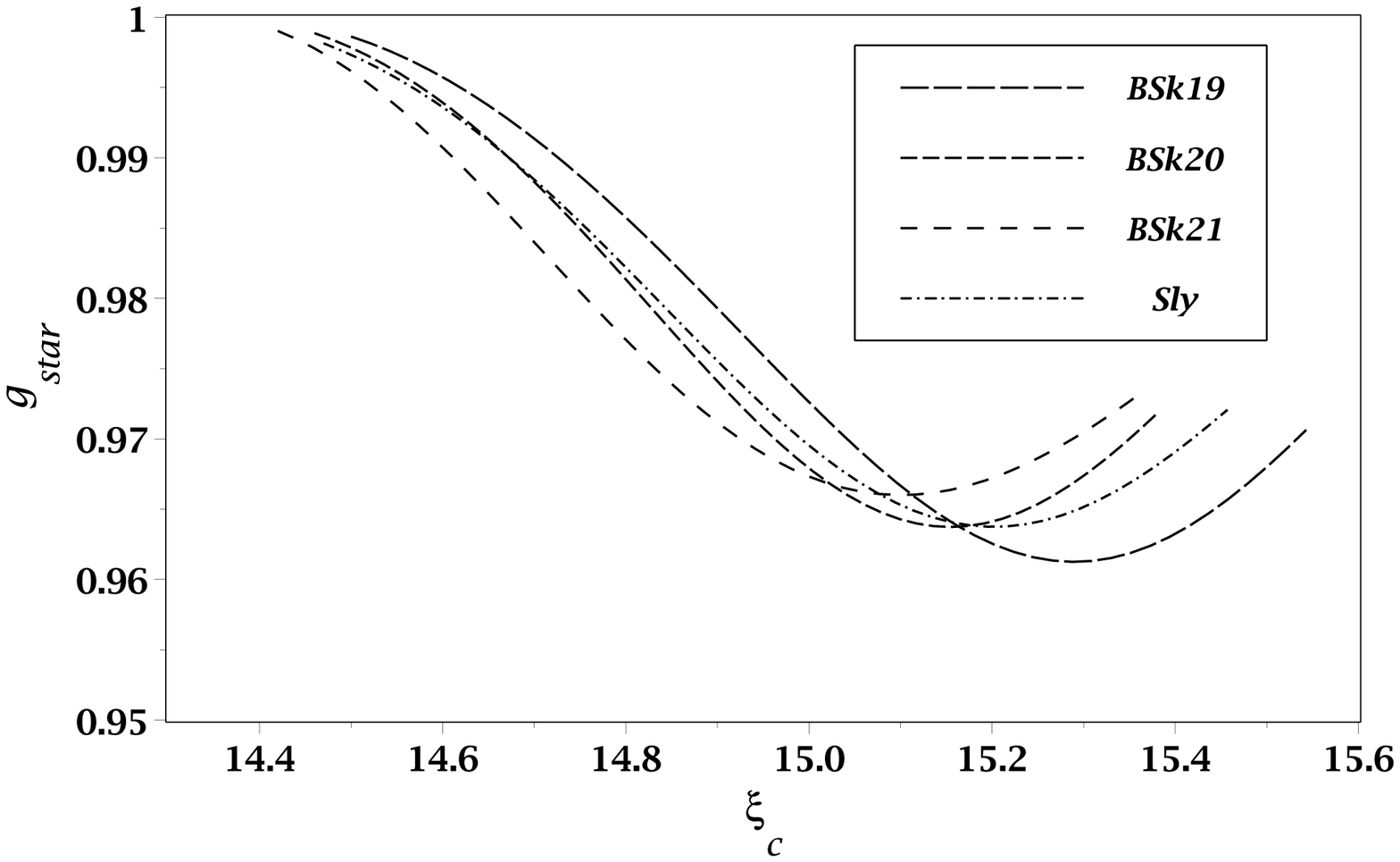}
\end{tabular}
\caption{Gravity intensity on the edge of the star as a function of the central density, for $d=10^{-20}$ (upper panel) and $d=10^{-22}$ (lower panel)}
\label{fig:gxi}
\end{figure}
\begin{figure}[t!]
\begin{tabular}{cc}
\includegraphics[width=0.95\columnwidth, angle=0]{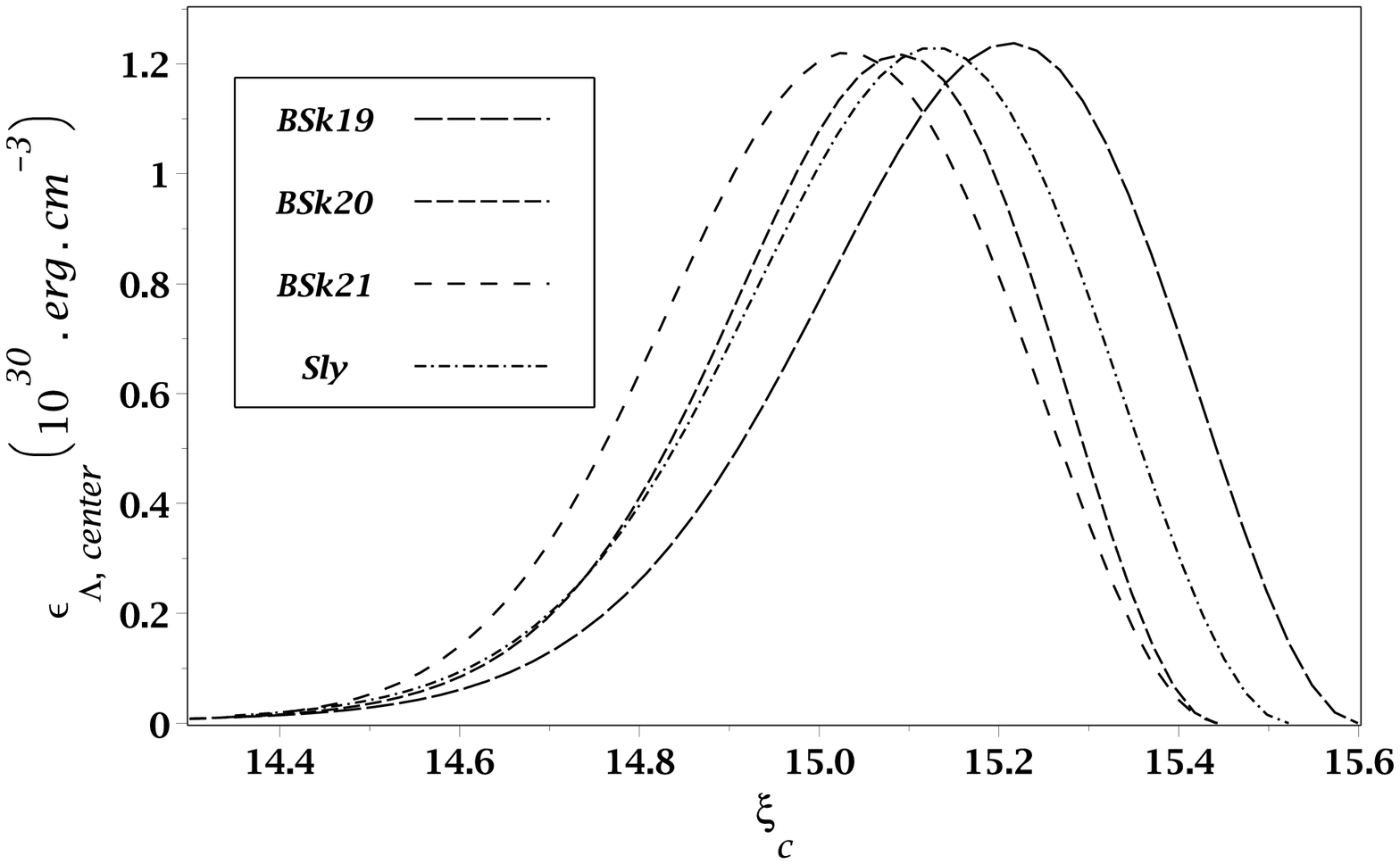} \\
\includegraphics[width=0.95\columnwidth, angle=0]{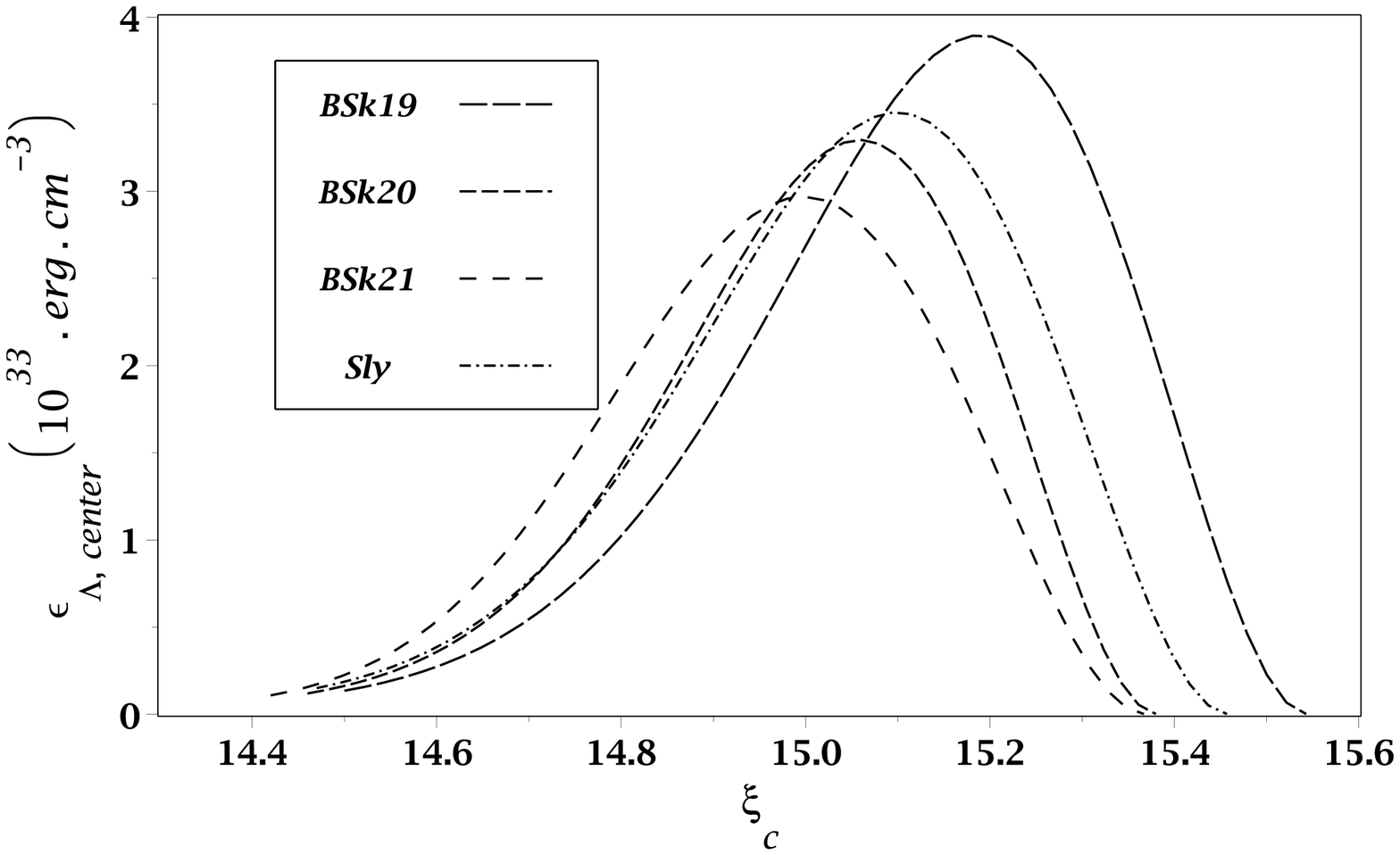}
\end{tabular}
\caption{Cosmological energy density $\epsilon_{\Lambda}$ in the center of the star as a function of the central density, for $d=10^{-20}$ (upper panel) and $d=10^{-22}$ (lower panel)}
\label{fig:elc}
\end{figure}
\begin{figure}[t!]
\begin{tabular}{cc}
\includegraphics[width=0.95\columnwidth, angle=0]{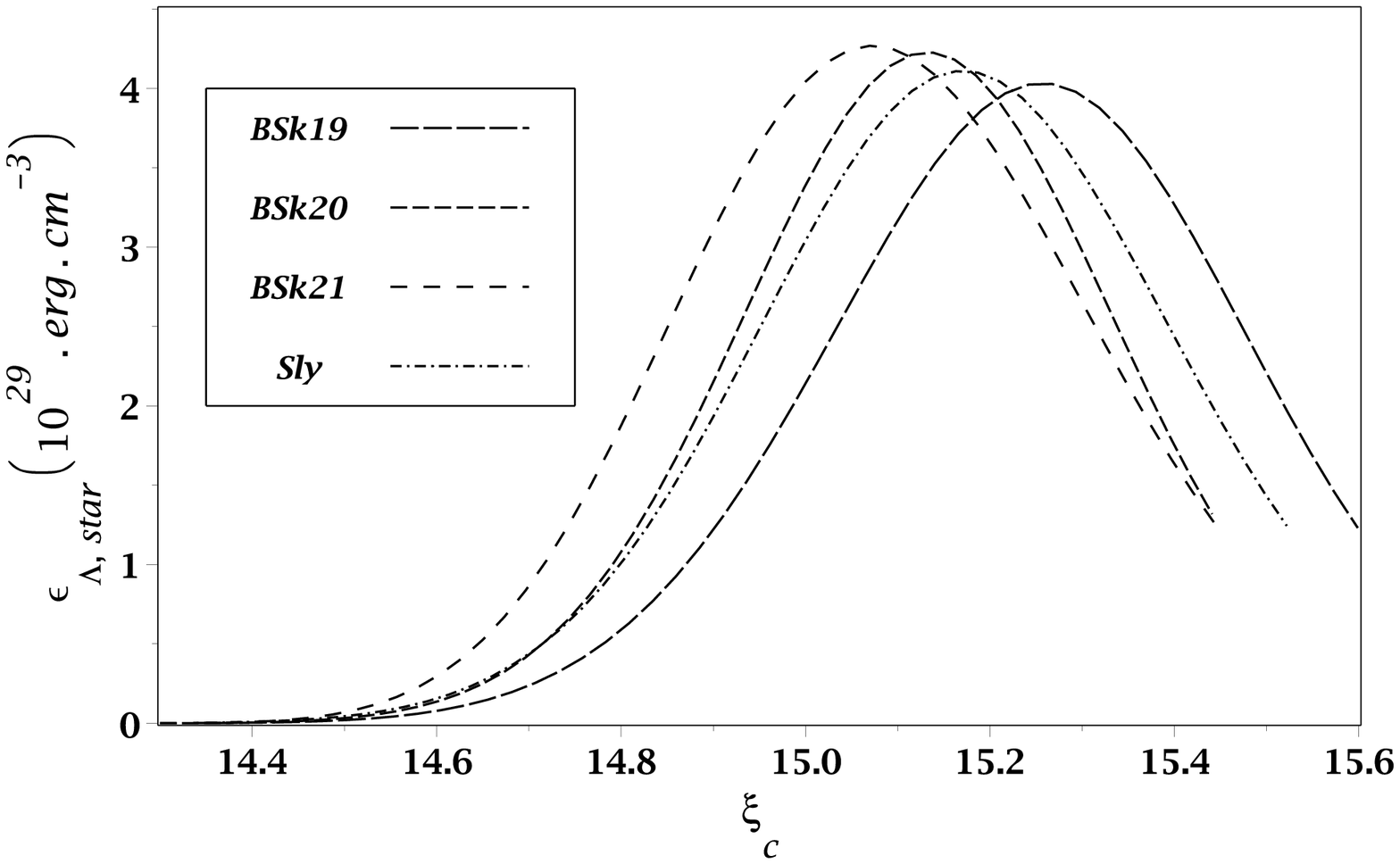} \\
\includegraphics[width=0.95\columnwidth, angle=0]{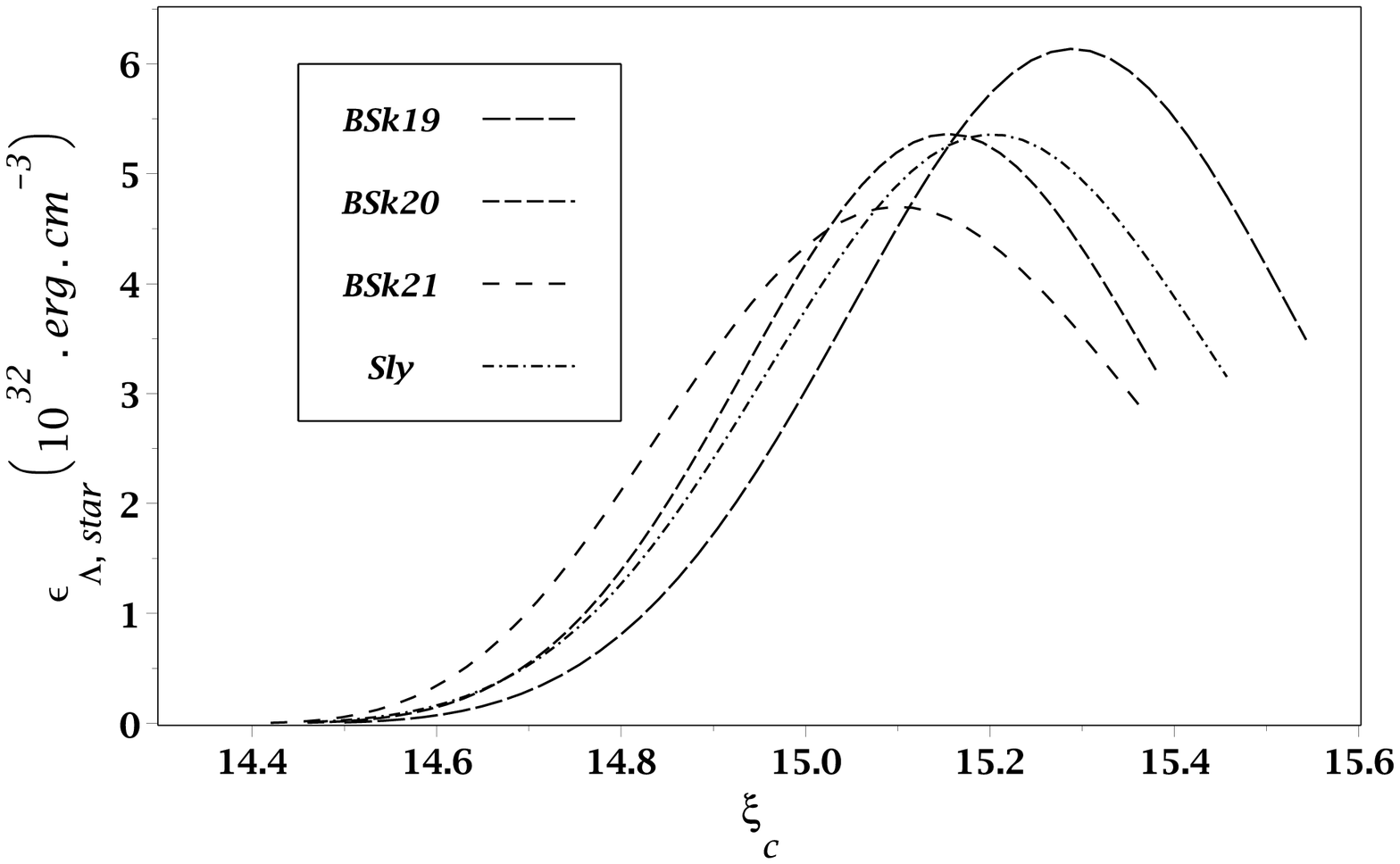}
\end{tabular}
\caption{Cosmological energy density $\epsilon_{\Lambda}$ on the edge of the star as a function of the central density, for $d=10^{-20}$ (upper panel) and $d=10^{-22}$ (lower panel)}
\label{fig:els}
\end{figure}
\begin{figure}[t!]
\begin{tabular}{cc}
\includegraphics[width=0.95\columnwidth, angle=0]{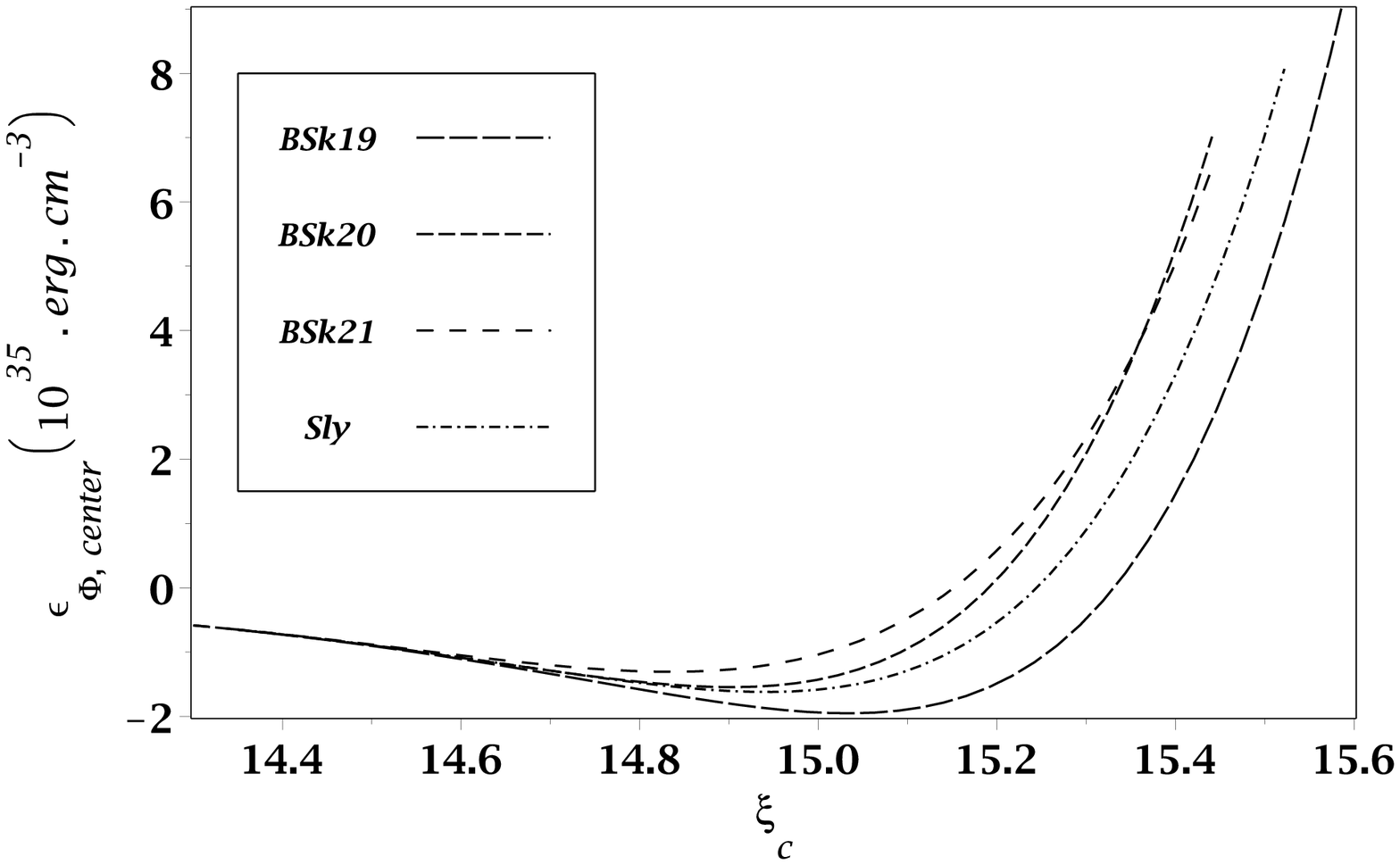} \\
\includegraphics[width=0.95\columnwidth, angle=0]{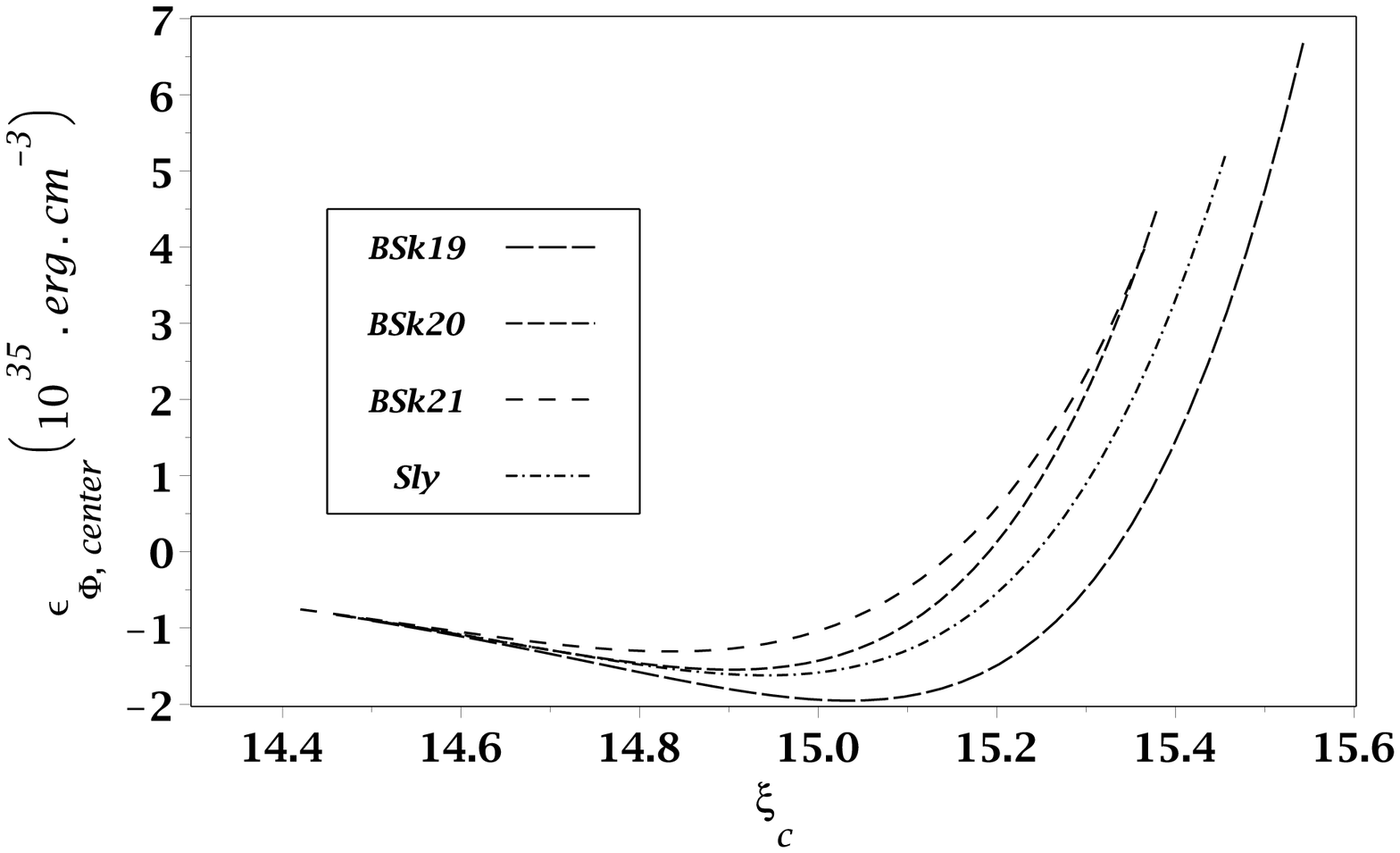}
\end{tabular}
\caption{Dilaton energy density $\epsilon_{\Phi}$ in the center of the star as a function of the central density, for $d=10^{-20}$ (upper panel) and $d=10^{-22}$ (lower panel)}
\label{fig:epc}
\end{figure}
\begin{figure}[t!]
\begin{tabular}{cc}
\includegraphics[width=0.95\columnwidth, angle=0]{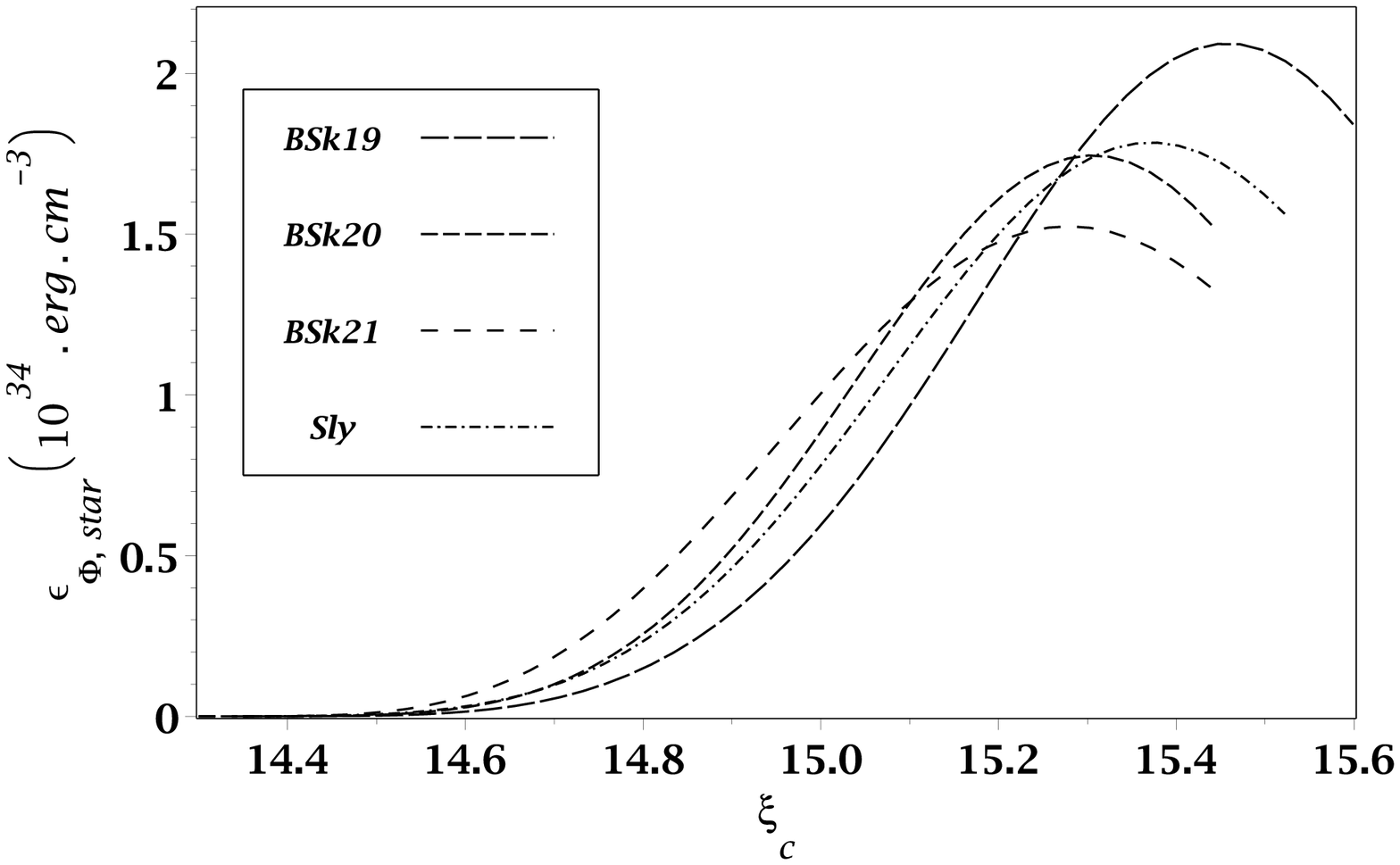} \\
\includegraphics[width=0.95\columnwidth, angle=0]{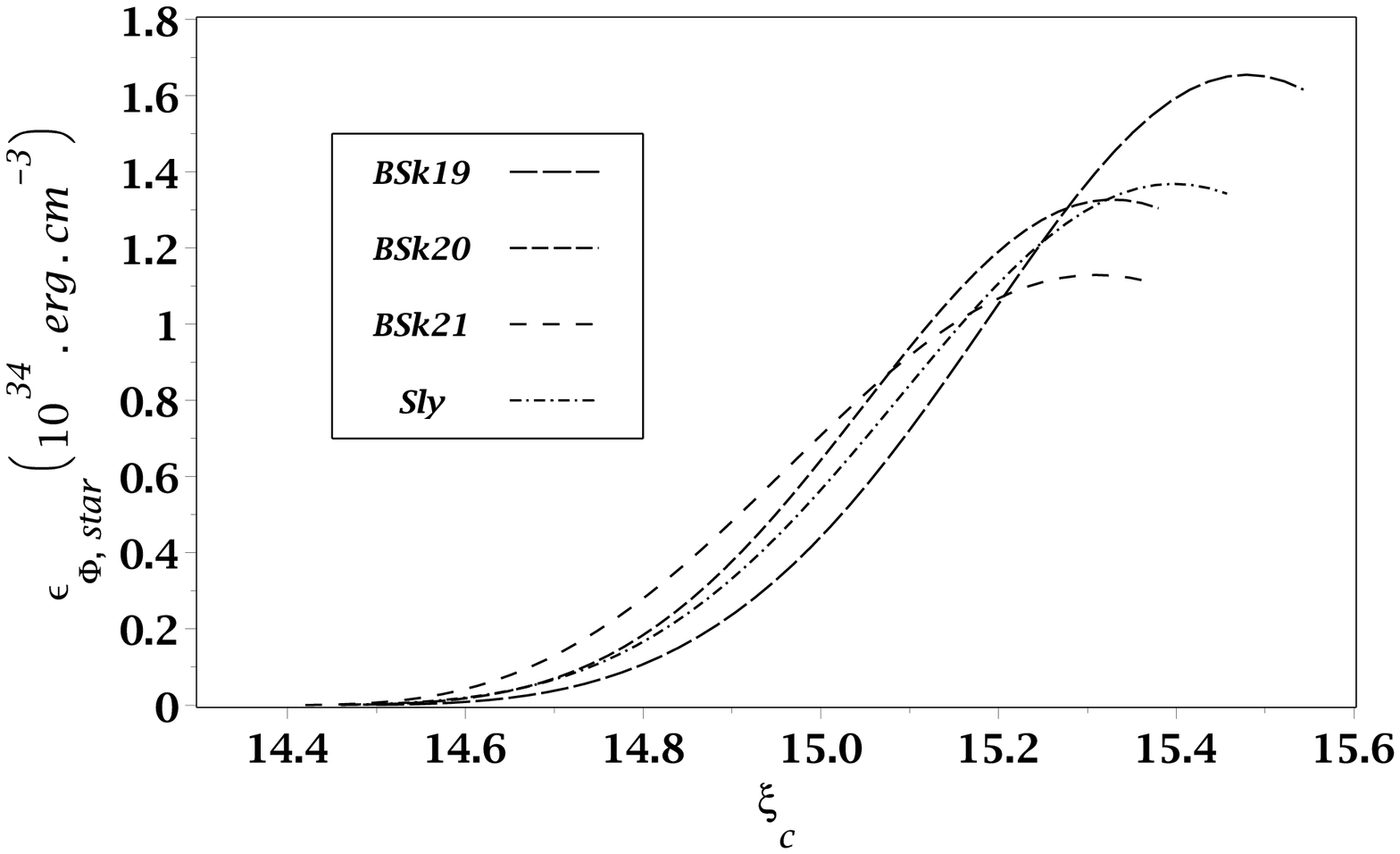}
\end{tabular}
\caption{Dilaton energy density $\epsilon_{\Phi}$ on the edge of the star as a function of the central density, for $d=10^{-20}$ (upper panel) and $d=10^{-22}$ (lower panel)}
\label{fig:eps}
\end{figure}

Another feature of the neutron star model in MDG is that it has a variational gravitational factor $G(\Phi)=G/\Phi$, instead of gravitational constant $G$. Although the dilaton field $\Phi$ has no direct interaction with ordinary matter, it influences the neutron star by changing the gravitational intensity, Fig.\ref{fig:gxi}. This effect justifies the name "dilaton" in MDG.

\section{Correspondence between f(R) gravity and MDG }
Study of the correspondence between MDG and f(R) theories is done in \citep{ref:4D} in the context of cosmology. It is known that the use of a cosmological potential $U(\Phi)$, of general form, is in general case only locally equivalent to f(R) theories of gravity \citep{ref:MDG2}. The use of the cosmological potential, as an alternative description of the nonlinear f(R) gravity is extremely useful. One is able to formulate simple and clear physical requirements for this potential. For example, to have a unique physical de Sitter vacuum in the theory, the function $U(\Phi)$ must have a unique positive minimum for positive values of the dilaton field $\Phi$. The potential must increase to infinity for $\Phi \rightarrow +0$, to avoid the nonphysical negative values of the dilaton field $\Phi$, i.e. the physical domain of antigravity.
\par
In our current study we use a simple one parameter potential \eqref{eqn:pot}, which corresponds to a nonpolynomial Lagrangian for the equivalent f(R) theory. It is described in parametric form by the equations $R=\frac{3}{4}d^{-2}(1/\Phi^{3}-\Phi)-4\Phi$ and $f=\frac{3}{8}d^{-2}(3/\Phi^{2}-\Phi^{2}-2)-2\Phi^{2}$, $\Phi \in (0,\infty)$.
\par
According to \cite{ref:MDG2} there is a Legendre transform between f(R) theories and MDG. But the two model are equivalent only under certain additional assumptions.They are equivalent only for certain class of potentials, like \eqref{eqn:pot}, called withholding potentials. We can translate the results from MDG to f(R) and visa versa, using the transformation from $U(\Phi)$ to f(R), and inverse transformation from f(R) to $U(\Phi)$, in the following parametric form:

\begin{subequations}\label{sub}
\begin{align}
f &{}=2(\Phi U,_{\Phi}(\Phi)-U(\Phi)) , \\
R &{}=2U,_{\Phi}, \Phi \in (0,\infty) , \\
U &{}=\frac{1}{2}(Rf,_{R}(R)-f(R)) , \\
\Phi &{}=f,_{R}, R \in (-\infty ,\infty ) .
\end{align}
\end{subequations}
In general this is only a local transformation, but not global, which could lead to a substantial different results concerning features of neutron star solutions.
\par
\begin{figure}[t!]
\begin{tabular}{cc}
\includegraphics[width=0.95\columnwidth, angle=0]{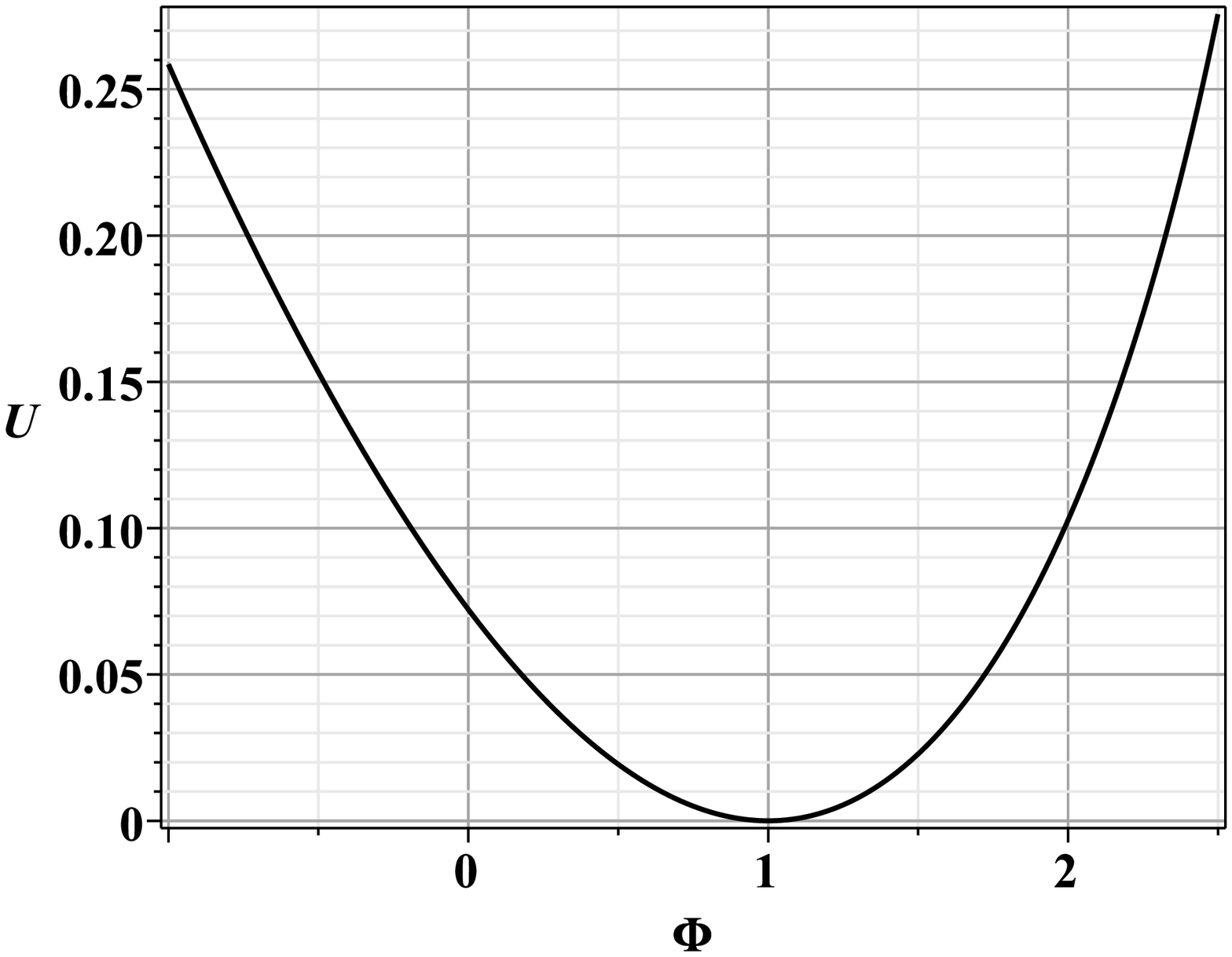} \\
\includegraphics[width=0.95\columnwidth, angle=0]{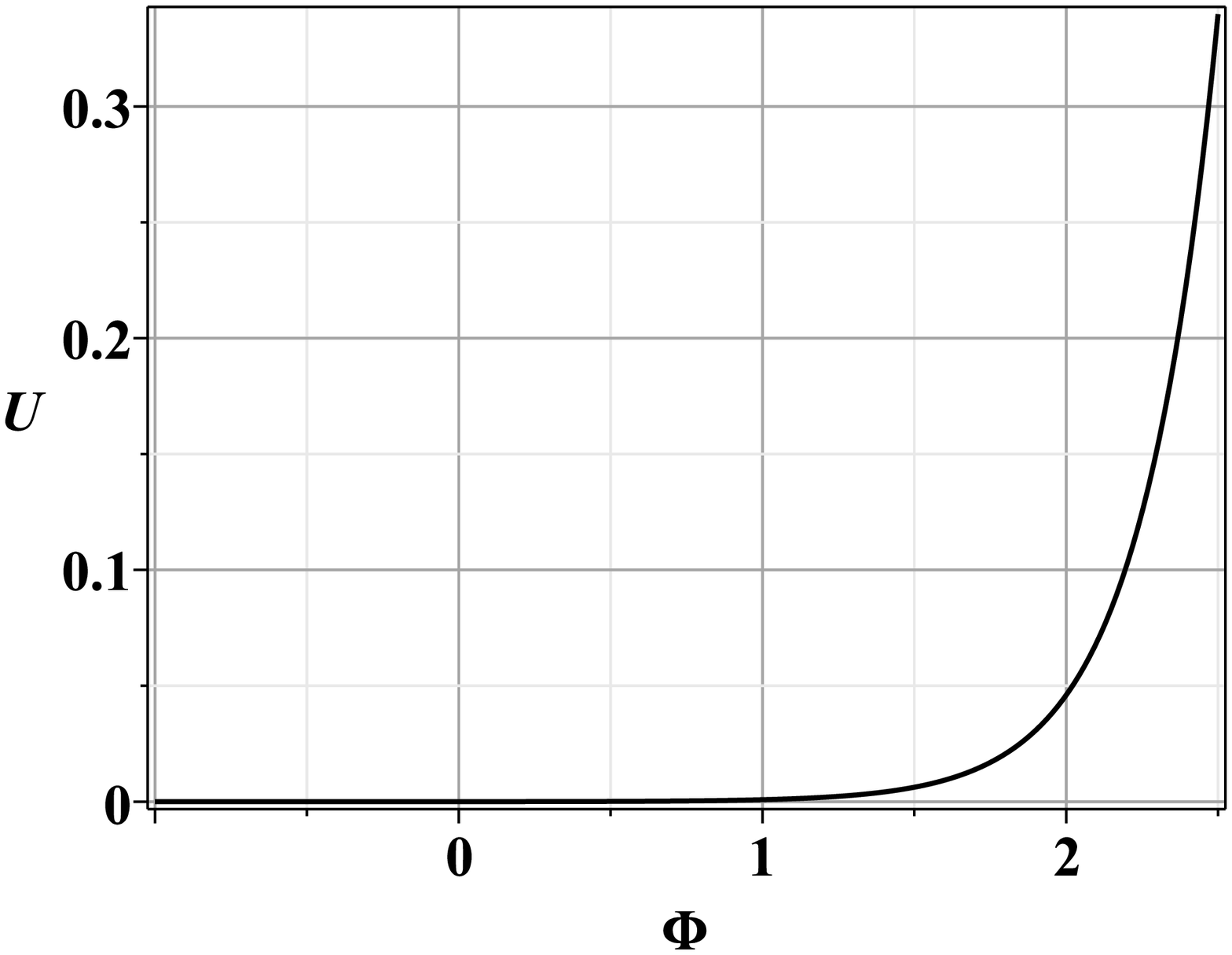}
\end{tabular}
\caption{The dilaton potential of f(R) function \eqref{subpot1}, $R_{0}=1, \beta=-1.5$ (upper panel).The dilaton potential of f(R) function \eqref{subpot2}, $\alpha=1, \beta=0.25, \mu=1$ (lower panel).}
\label{fig:pot1}
\end{figure}
\begin{figure}[t!]
\begin{tabular}{cc}
\includegraphics[width=0.95\columnwidth, angle=0]{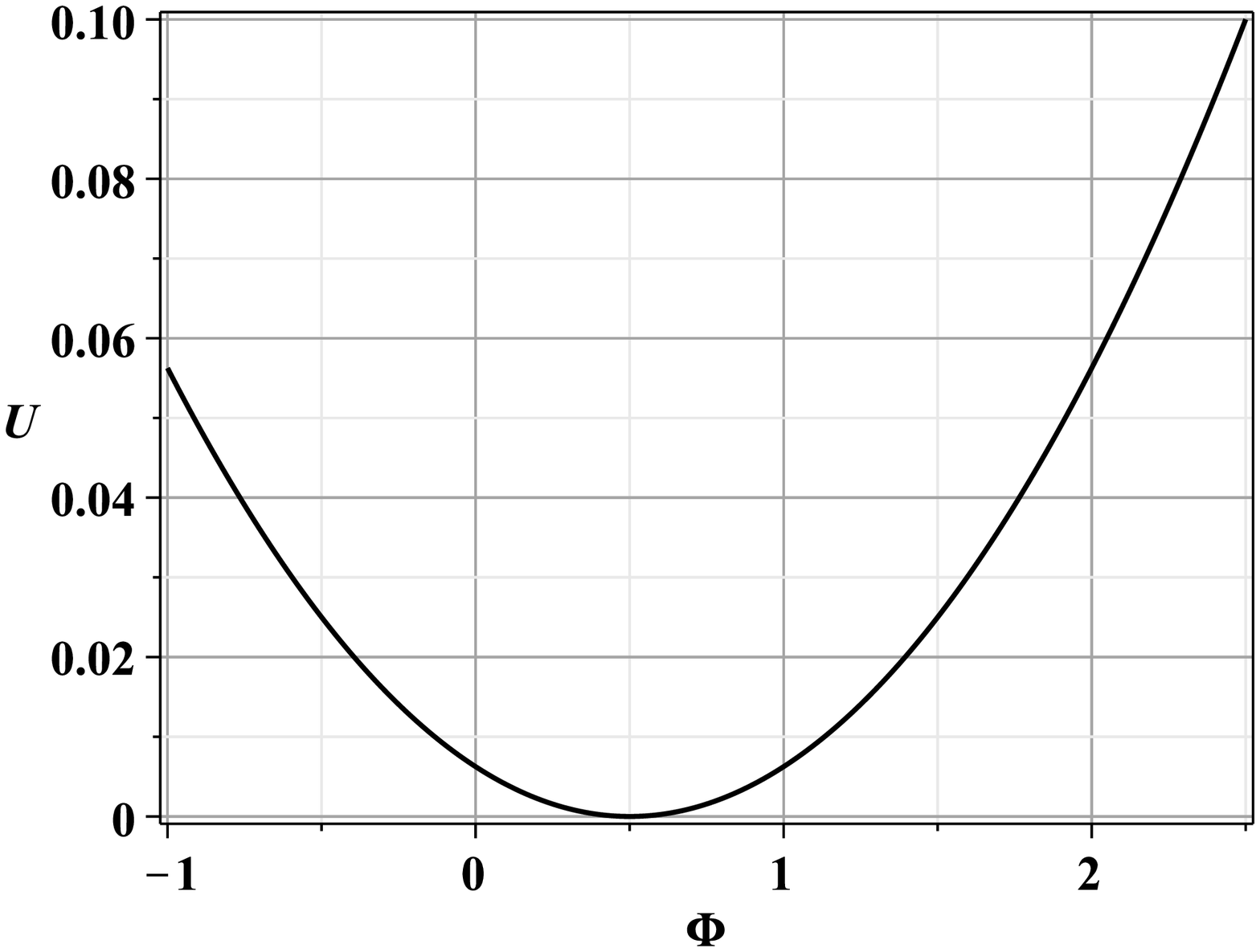} \\
\includegraphics[width=0.95\columnwidth, angle=0]{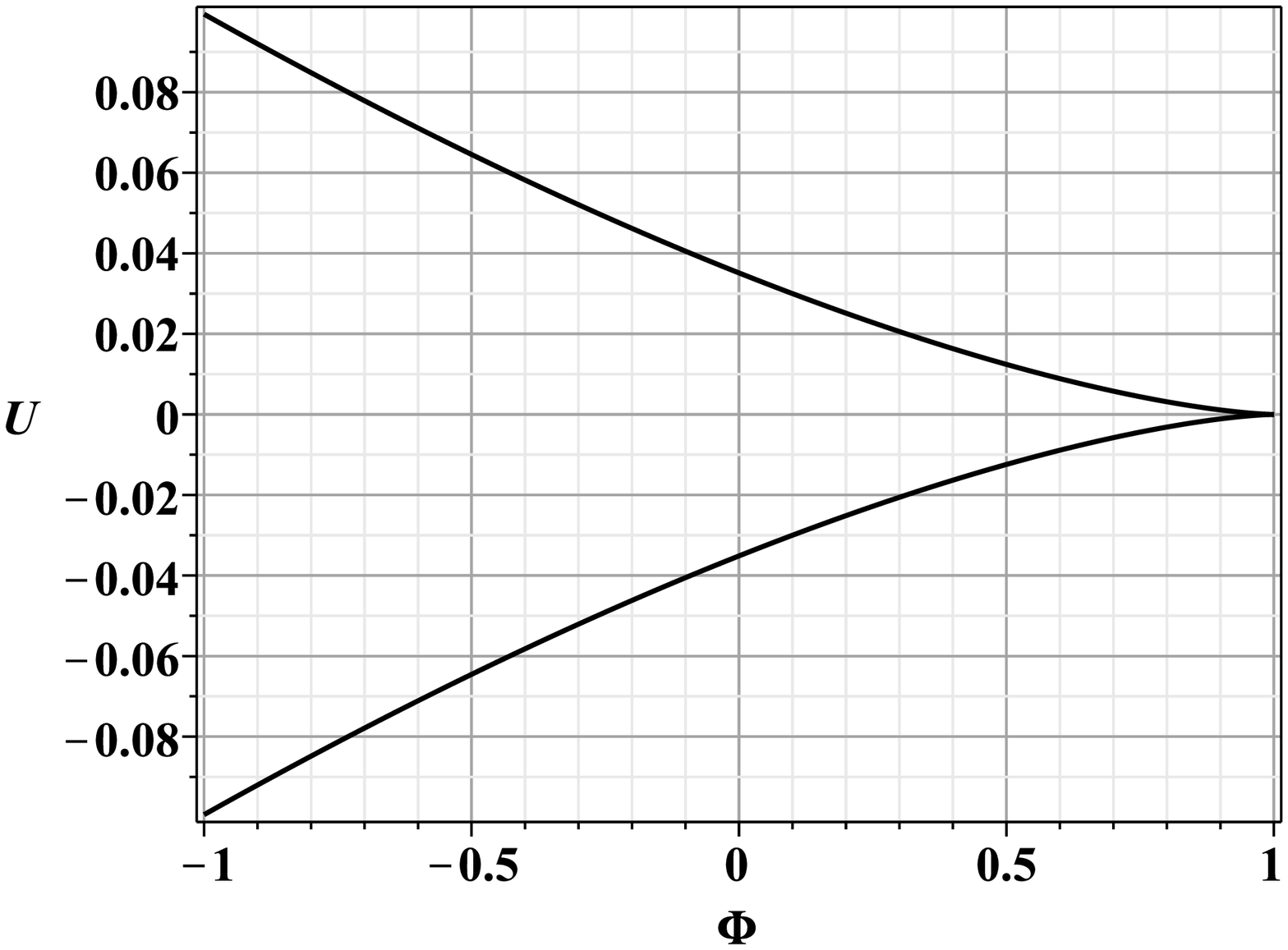}
\end{tabular}
\caption{The dilaton potential of f(R) function \eqref{subpot3}, $\alpha=-0.5, \gamma=-10$ (upper panel).The dilaton potential of f(R) function \eqref{subpot4}, $\beta=-40, \gamma=-0.45$ (lower panel).}
\label{fig:pot2}
\end{figure}

Lets take, for example, models with
\begin{subequations}\label{models}
\begin{align}
f(R) &{}=R+\beta R(exp(-R/R_{0})-1) , \label{model1} \\
f(R) &{}=R+\alpha R(1+\beta ln(R/ \mu)) , \label{model2} \\
f(R) &{}=R+\alpha R(1+\gamma R) , \label{model3} \\
f(R) &{}=R+\beta R^{3} ,\label{model4}
\end{align}
\end{subequations}
where $R_{0}$ is constant, $\alpha , \beta , \gamma, \mu$ are parameters, different for the every model. Using the transformations \eqref{sub} one can evaluate the corresponding potentials $U(\Phi)$.

\begin{subequations}\label{potentials}
\begin{align}
U(\Phi) &{}=-\frac{1}{2} \frac{(\Phi +\beta -1)((eW(\Phi + \beta -1)/\beta)-1)^{2}}{W(\Phi + \beta -1)/\beta)} , \label{subpot1} \\
U(\Phi) &{}=-\frac{\alpha \beta \mu}{2} exp \left(\frac{\Phi -\alpha \beta -\alpha-1}{\alpha \beta}\right) , \label{subpot2} \\
U(\Phi) &{}=\frac{1}{8} \frac{(\Phi -\alpha -1)^2}{\alpha \gamma} , \label{subpot3} \\
U(\Phi) &{}=\pm \frac{1}{9} \frac{\sqrt{3}(\beta (\Phi -1))^{3/2}}{\beta^{2}} , \label{subpot4}
\end{align}
\end{subequations}
where W(x) is Lambert W-function. The potentials are plotted on Fig.\ref{fig:pot1} and on Fig.\ref{fig:pot2}. All models lead to potential that allow values $\Phi<0$ and \eqref{subpot4} leads to a multivalued function $U(\Phi)$. Two models \eqref{model1} and \eqref{model3}  lead to convex functions, which is a requirement for the withholding property of the dilaton potential, but they are not in the physical domain in the corresponding model of MDG.
\par
The f(R) models \eqref{models} do not satisfy the withholding property of the corresponding dilaton potentials, which therefore are unphysical, according to MDG criteria, but if restricted to certain subregions of the dilaton field they could have partially uncontroversial physical meaning. Such restrictions will not work in quantum theory when under-barrier quantum transitions are allowed. Another problem is that one cannot choose natural justified and independent criteria for restraining the f(R) function in that subregion.

\section{Discussion}
This paper presents more a extensive research on the topic of neutron stars in MDG. It is based on the previous research on the topic \citep{ref:MDG3,ref:MDG4,ref:MDG5,ref:MDG6,ref:MDG7}. All the results are obtained using the analytical representation of some of the most physically meaningful EOS: SLy, BSk19, BSk20, BSk21, and are qualitatively in good accordance with the previous results on static spherically symmetric stars in MDG \citep{ref:MDG3,ref:MDG5,ref:MDG6}. Different dilaton masses are used in the present paper, which give us insight into how various physical quantities depend on the dilaton field. The calculation of the cosmological ($p_{\Lambda},\epsilon_{\Lambda}$) and dilaton ($p_{\Phi},\epsilon_{\Phi}$) variables shows the dependence on the matter initial conditions and helps us understand their behavior and their interpretation as possible candidates for dark energy and dark matter.
\par
The results are also in accordance with the latest observational data for the maximum mass of a neutron star. The important role of the dilasphere is confirmed. It is a sort of dark matter halo and carries about $15\div 30\%$ of the mass of the entire object. It appears for all EOS, dilaton masses and initial conditions. 
\par
This work and previous ones open new possibilities for research in the MDG model. A model of rotating stars in MDG may lead to asymmetry of the matter and dilaton field configurations. The research of different stellar objects and gravitationally interacting systems may also lead to new and interesting results.

\section{Contribution of the authors}
K.M. was the operating person in all numerical calculations in the present paper. He obtained all the results for EOS: SLy, BSk19, BSk20, BSk21. He made all figures and wrote the initial text of the article.\\
P.F. supervised the project, derived the basic equations and boundary conditions described in the Introduction and Section 2. He wrote the computer program for numerical calculations using the Maple grid parallel programing, used also in \citep{ref:MDG6}.\\
Both the authors discussed all the obtained new results and physical interpretation together. 
\par
\acknowledgments
Kalin Marinov wants to express gratitude to the "Program for career development of young scientists, Bulgarian Academy of Sciences" project "Extended theories of gravity and their application to physics of compact stars " № DFNP - 51 / 21.04.2016.\\
Plamen Fiziev is deeply indebted to the Directorate of the Laboratory of Theoretical Physics, JINR, Dubna, for the good working conditions and support.\\
He also owes gratitude to Alexei Starobinsky, Salvatore Capozziello, Sergei Odintsov, Mariafelicia De Laurentis, Alexander Zacharov, Luciano Rezzolla, Valeria Ferrari, Pawel Haensel, Alexander Potekhin and Fiorella Burgio for the stimulating discussions on different topics of the present publication.\\
Special thanks to Kazim Yavuz Ek\c{s}i for providing the numerical EOS data.\\
The authors are thankful to the unknown referee for the useful comments of the initial version o the paper. Especially, the present new Section 4 is our answer to the question raised by the referee.\\ 
This research was supported in part by the Foundation for Theoretical and Computational Physics and Astrophysics and by Bulgarian Nuclear Regulatory Agency Grants for 2014, 2015 and 2016 as well as by "NewCompStar", COST Action MP 1304.

\end{document}